\documentclass[manuscript]{acmart}

\usepackage{xspace}
\usepackage{hyperref}
\usepackage[capitalise]{cleveref}
\usepackage{subcaption}
\usepackage{enumitem}
\usepackage{scratch3}

\newcommand{\setscratchA}{\setscratch{scale=0.7}\raisebox{1.2pt}}

\newcommand{\setscratchB}{\setscratch{scale=0.53, y sepinf=3pt, y sepsup=0pt}\raisebox{0pt}}

\newcommand{\setscratchC}{\setscratch{scale=0.6, y sepinf=3pt, y sepsup=0pt}\raisebox{1.2pt}}

\renewcommand{\paragraph}[1]{\vspace{0.08em}\noindent {\bf #1}}
\let\oldtextsf\textsf
\renewcommand{\textsf}[1]{{\small \oldtextsf{#1}}}

\newcommand{\VM}{virtual machine\xspace}

\newcommand{\inline}[1]{%
	\includegraphics[height=0.9\baselineskip]{#1}
}

\newcommand{\summary}[2]{
	\vspace{0.4em}
	\noindent
	\colorbox{gray!20}{%
		\parbox{.97\linewidth}{%
			\textbf{\textsf{Summary (\textit{#1})}}
			#2
		}%
	}\vspace{0.4em}%
}%

\newcommand\definetool[2]{\newcommand{#1}{{\textsc{#2}}\xspace}}
\definetool{\Scratch}{Scratch}
\definetool{\LitterBox}{LitterBox}
\definetool{\drscratch}{Dr. Scratch}
\definetool{\qualityhound}{Qualityhound}
\definetool{\hairball}{Hairball}
\definetool{\itch}{Itch}
\definetool{\Whisker}{Whisker}
\definetool{\bastet}{Bastet}
\definetool{\Nuzzlebug}{NuzzleBug}
\definetool{\Catnip}{Catnip}
\definetool{\Poke}{Poke}

\setcopyright{acmlicensed}
\copyrightyear{2024}
\acmYear{2024}
\acmDOI{XXXXXXX.XXXXXXX}

\acmConference[Koli Calling 2024]{24th Koli Calling International Conference on Computing Education Research Proceedings}{November 2024}{Koli, Finland}
\acmISBN{978-1-4503-XXXX-X/18/06}

\begin{document}

\title{A Block-Based Testing Framework for Scratch}

\author{Patric Feldmeier}
\authornote{Authors are listed in alphabetical order.}
\email{patric.feldmeier@uni-passau.de}
\affiliation{%
	\institution{University of Passau}
	\country{Germany}
}

\author{Gordon Fraser}
\authornotemark[1]
\email{gordon.fraser@uni-passau.de}
\affiliation{%
	\institution{University of Passau}
	\country{Germany}
}

\author{Ute Heuer}
\authornotemark[1]
\email{ute.heuer@uni-passau.de}
\affiliation{%
	\institution{University of Passau}
	\country{Germany}
}

\author{Florian Obermüller}
\authornotemark[1]
\email{obermuel@fim.uni-passau.de}
\affiliation{%
	\institution{University of Passau}
	\country{Germany}
}

\author{Siegfried Steckenbiller}
\authornotemark[1]
\email{steckenbiller@fim.uni-passau.de}
\affiliation{%
\institution{University of Passau}
\country{Germany}
}

\begin{abstract}
  Block-based programming environments like \Scratch are
  widely used in introductory programming courses. They
  facilitate learning pivotal programming concepts by
  eliminating syntactical errors, but logical errors that
  break the desired program behaviour are nevertheless
  possible. Finding such errors requires \emph{testing}, i.e.,
  running the program and checking its behaviour. In many
  programming environments, this step can be automated by
  providing executable tests as code; in \Scratch, testing can
  only be done manually by invoking events through user input and
  observing the rendered stage. While this is arguably
  sufficient for learners, the lack of automated testing may
  be inhibitive for teachers wishing to provide feedback on
  their students' solutions.
  In order to address this issue, we introduce a new category
  of blocks in \Scratch that enables the creation of automated
  tests. With these blocks, students and teachers alike can create
  tests and receive feedback directly within the \Scratch environment
  using familiar block-based programming logic. To facilitate the creation and to enable batch processing sets of student solutions, 
  we extend the \Scratch user interface with an accompanying
  test interface.
  %
  We evaluated this testing framework with 28 teachers who created
  tests for a popular \Scratch game and subsequently used these tests
  to assess and provide feedback on student implementations. An
  overall accuracy of 0.93 of teachers' tests compared to manually
  evaluating the functionality of 21 student solutions demonstrates
  that teachers are able to create and effectively use tests. A
  subsequent survey confirms that teachers consider the
  block-based test approach useful.
\end{abstract}


\maketitle

\newcommand{\meanAccuracyFeedbackSample}{0.92\xspace}
\newcommand{\meanAccuracyTestFeedback}{0.96\xspace}
\newcommand{\meanAccuracyTestSample}{0.93\xspace}
\newcommand{\totalChanges}{150\xspace}
\newcommand{\totalChangesWorse}{77\xspace}
\newcommand{\totalChangesBetter}{73\xspace}
\newcommand{\totalPassToFailCorrect}{23\xspace}
\newcommand{\totalFailToPassCorrect}{50\xspace}
\newcommand{\totalPassToFailWrong}{45\xspace}
\newcommand{\totalFailToPassWrong}{32\xspace}
\newcommand{\meanBetter}{0.42\xspace}
\newcommand{\meanWorse}{0.58\xspace}
\newcommand{\meanChanges}{4.0\xspace}
\newcommand{\meanPassToFailCorrect}{0.0\xspace}
\newcommand{\meanFailToPassCorrect}{0.5\xspace}
\newcommand{\meanFailToPassWrong}{1.0\xspace}
\newcommand{\meanPassToFailWrong}{2.0\xspace}

\section{Introduction}
\label{sec:introduction}

\Scratch makes writing code easy: There is neither a need to memorise
programming commands nor programming syntax, and programs are created
by simply dragging and dropping the available programming blocks to
assemble scripts, which are syntactically correct if the blocks'
shapes interlock. However, even a syntactically correct program can be
functionally broken. Determining whether or not a program implements
the desired functionality requires testing, i.e., running the program
and validating the resulting behaviour.
In other programming environments, this process can be automated by
writing code that interacts with the program and checks the resulting
behaviour. Running such test code provides feedback on functionality
automatically, quickly, and repeatedly.
In \Scratch, however, testing has to be done entirely manually by
interacting with the program. While this may be
acceptable for learners whose focus is on being creative, it inhibits
automated analyses or teachers who need to inspect programs to provide
feedback.

In recognition of this gap, research prototypes for testing \Scratch
programs have previously been
proposed~\cite{stahlbauer2019testing,johnson2016itch}, demonstrating
just how useful tests can be for \Scratch, for example by automating
tedious tasks such as assessing student
programs~\cite{stahlbauer2019testing} or providing
feedback~\cite{obermueller2023tutorials}. However, these testing
frameworks usually require tests to be written in other languages, such
as JavaScript or Python, and require additional tooling beyond the
\Scratch interface for execution. This, unfortunately, represents a
fundamental inhibitor hampering adoption by teachers.

\begin{figure}[!t]
  \centering
  \begin{subfigure}[c]{0.44\textwidth}
    \includegraphics[width=0.95\textwidth]{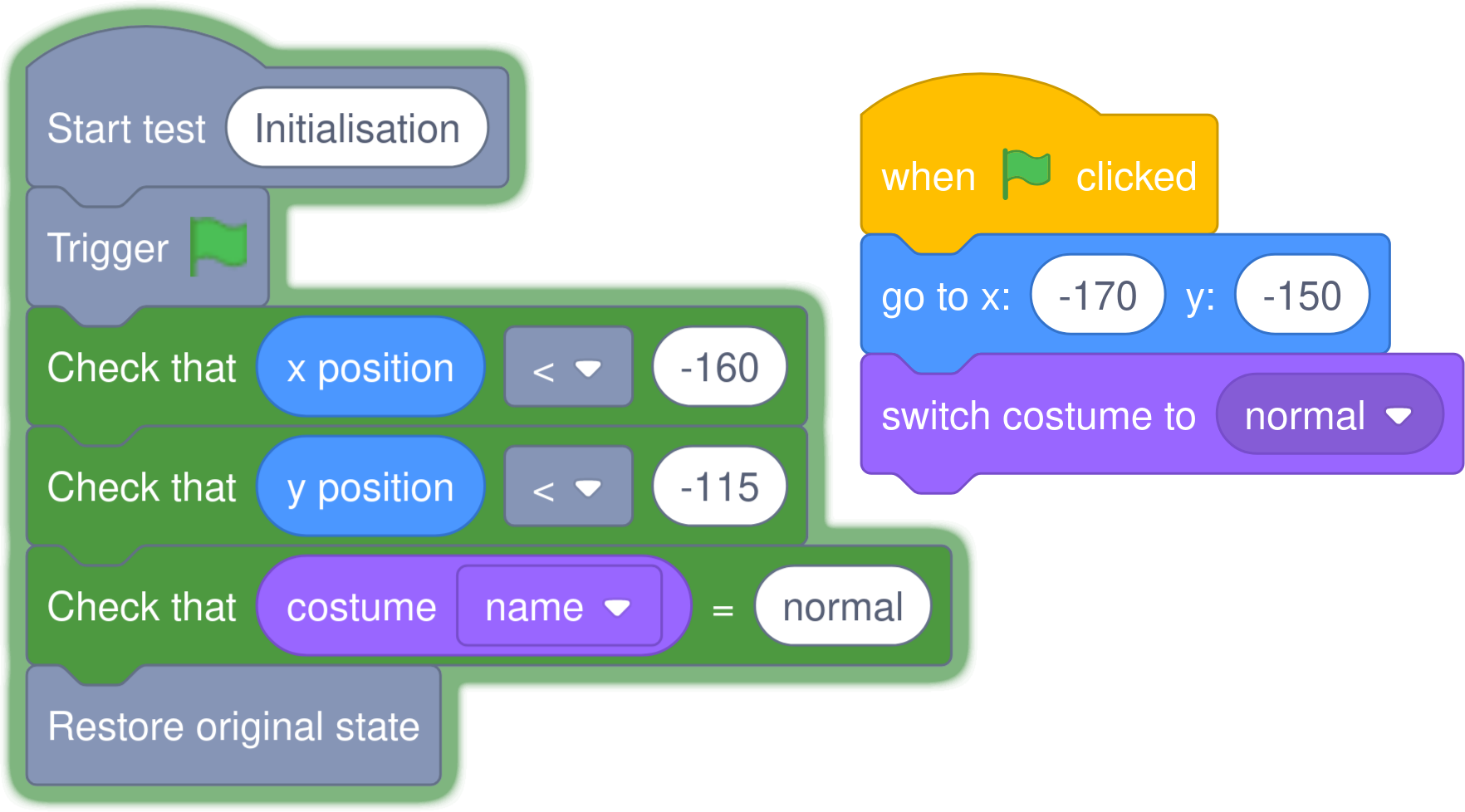}
    \subcaption{\label{fig:example_pass}The correct initialisation script makes the test pass.}
  \end{subfigure}
  \begin{subfigure}[c]{0.44\textwidth}
    \includegraphics[width=0.95\textwidth]{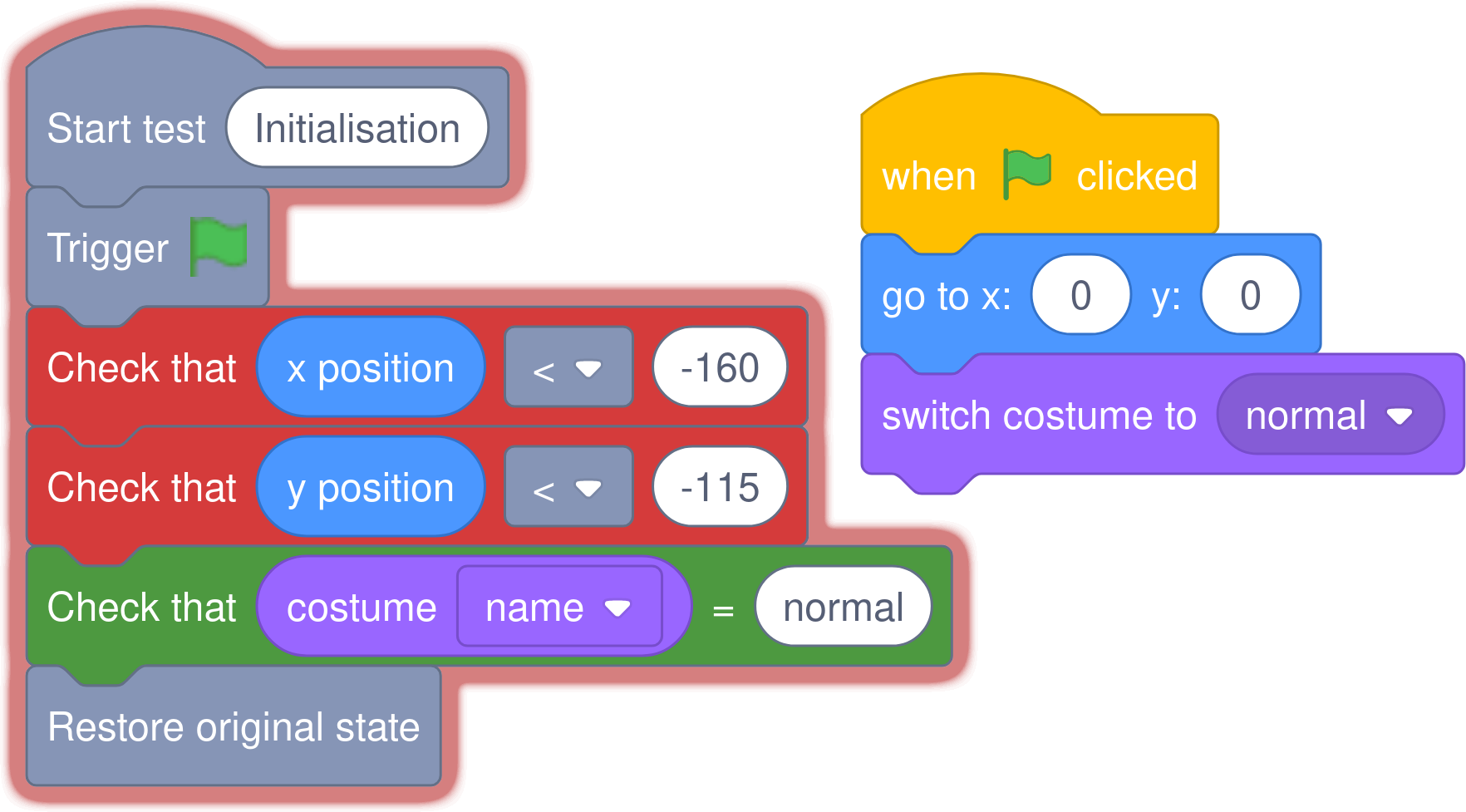}
    \subcaption{\label{fig:example_fail}The faulty initialisation script makes the test fail.}
  \end{subfigure}
	\vspace{-0.5em}
    \caption{A correct and a faulty variant of an initialisation script are tested.}
    \label{fig:example}
    \vspace{-1em}
\end{figure}

In this paper we therefore introduce a \Scratch extension for creating
and executing tests directly within \Scratch.  \Cref{fig:example_pass}
shows a program that sets the position of a sprite to (-170, -150) and
changes the costume to \emph{normal}, together with a corresponding
block-based test.
When executed, the test triggers program execution through a green
flag event, subsequently checking the sprite's position and costume.
The green highlighting of the test in \cref{fig:example_pass}
indicates that the program exhibited the expected behaviour. On the
other hand, the script in \cref{fig:example_fail} incorrectly sets the
position to (0, 0), causing the test to fail as the position does not
reach the expected state, visualised by red highlighting. Creating and
running such tests is easy and provides immediate feedback within the
Scratch editor, allowing students to self-evaluate their
implementation with teacher-provided tests automatically. As tests are just further
executable blocks, both teachers and learners are able to use these
tests to receive automated feedback on a \Scratch project through the
colour-coded highlighting and error messages in the control interface.
In detail, the contributions of this paper are as follows:
\begin{itemize}
\item Block-based testing: We extend \Scratch with test-related blocks
  for creating tests, and modify \Scratch to support their execution
  and evaluation.
  
\item Test control interface: We extend the \Scratch user interface to
  control test execution and help interpret test results. Tests can
  be run individually or collectively, on individual programs or on
  batches of student solutions.
\item Empirical evaluation: We evaluate the block-based testing
  framework with 28 teachers tasked to create tests and to use these
  tests when assessing student code.
\end{itemize}
Our study demonstrates that teachers can write effective tests
suitable for assessing student solutions, 
 and their feedback is encouraging. Our
open source implementation is available for future research and
education on testing in \Scratch.


\section{The Scratch Programming Environment}
\label{sec:background}

\Scratch\cite{maloney2010} is a block-based programming environment
that aims to make programming more accessible for novices, and it is one
of the most popular~\cite{mcgill2020} programming environments for
learning.
\Scratch favours exploration over recall by displaying all blocks in a
toolbox, with programs being built by visually arranging these blocks
(which represent statements and expressions) into scripts. Different
block shapes demonstrate which combinations are valid. For example,
boolean expressions are represented as diamond blocks
\setscratchA{\boolsensing{bool}}, while
other expressions are rounded blocks \setscratchA{\ovalsensing{X}}, and wherever
these can be inserted, there are gaps of the same shape. Statements
can be stacked, such that it is only possible to create syntactically
valid programs.
Many statements represent high-level actions of \emph{sprites}
interacting on a \emph{stage}, which, together with simple media
integration~\cite{bau2017}, makes it easy to combine blocks to result
in game-like programs.

As the underlying engine, the \Scratch \VM is responsible for executing
block-based programs, maintaining a representation of a program's
current state and handling the event-driven nature of \Scratch by
responding to user inputs. The execution of programs is managed by a
sequencing function that mimics the parallel execution of scripts
through a threading model, where each script is associated with a
separate thread. Whenever a script is activated, e.g., due to user
input, the corresponding thread is added to the set of currently
active threads. During program execution, each active thread is given
permission by the sequencing function to execute the
corresponding script until either its end is reached or
 execution halting blocks that cause the thread to yield are
encountered, such as the
\setscratchC{\begin{scratch}\blockcontrol{Wait for \ovalnum{1} seconds}\end{scratch}} or
\setscratchC{\begin{scratch}\blockcontrol{wait until \boolempty[1em]}\end{scratch}}
blocks~\cite{deiner2023automated}.
If the execution of a script was paused due to an
execution halting block, its execution is resumed after the sequencer
has processed all other currently active threads.

The different block shapes in \Scratch prevent syntactical errors so
learners can focus on what the program should do, but it can
nevertheless be challenging to achieve the intended behaviour,
resulting in code containing bugs. While the process of finding and
fixing bugs is a crucial part of learning to code, tools can offer
support to learners as well as teachers.
Misconceptions of learners often manifest in bugs that follow
recurring bug patterns (code idioms that are likely to be
defects~\cite{hovemeyer2004}). Bug patterns for \Scratch have been
demonstrated to occur frequently~\cite{fraedrich2020} and can be
detected automatically by linters~\cite{fraser2021litterbox}.
Such linters perform static analyses of programs, thus applying
general criteria that are suitable for any \Scratch project. However,
checking specific functionality, such as whether a sprite is placed at
a given location, is task-specific and also requires running the
program to test its behaviour.

Existing approaches for checking the functionality of \Scratch
programs include automated testing tools like
\itch~\cite{johnson2016itch} and
\Whisker~\cite{stahlbauer2019testing}, or model checking tools like
\bastet~\cite{stahlbauer2020}. \Whisker has been shown to be useful
for automated grading~\cite{stahlbauer2019testing}, to support
next-step hint generation~\cite{fein2022,obermueller2021hinttest}, and
for interactive \Scratch tutorial
systems~\cite{obermueller2023tutorials}.
%
To utilise such approaches, educators need to manually create test
cases in the syntax required by these tools (e.g., JavaScript), for
example by describing the order in which inputs should be sent to the
program under test and what the expected program response should be,
or they have to formally specify expected behaviour. However, this
poses a challenge for educators not well versed in the respective
syntax demanded by these tools.

In this paper we therefore investigate the idea of integrating
testing directly into the \Scratch user interface. Concurrently to our
work, there have been independent proposals of similar ideas: In a
recent master thesis~\cite{poke}, the \Poke extension of \Scratch
proposed a new category of blocks that enables the creation of test cases
similar to the blocks we propose in this paper. An alternative
proposal~\cite{nurue2024} consists of an individual new test block
that demonstrates the concept of creating automated tests through custom \Scratch blocks.
Our approach is the first to offer a fully integrated, block-based
test framework that makes automated and systematic testing of \Scratch
projects available without external tools to anyone being familiar with the
\Scratch programming language.




\section{Test-Related Blocks}
\label{sec:approach}

As one of the core contributions of this work, we introduce a new
category of blocks that provide test-related functionality. Testing
blocks encompass a diverse range of capabilities, organised into four
categories: \emph{control blocks}, which manage the execution of
tests~(\cref{sec:control}); \emph{event and input blocks}, which
simulate events and user inputs to trigger corresponding event
handlers~(\cref{sec:trigger}); \emph{assertion blocks}, which evaluate
conditions and determine the test result~(\cref{sec:assertion}); and
\emph{reporter blocks}, which facilitate writing tests by providing
access to frequently required sprite attributes~(\cref{sec:reporter}).

\subsection{Control Blocks}
\label{sec:control}

\begin{figure}[!tbp]
    \centering
    \subfloat[\label{fig:controlBlocks}\centering Control blocks for managing 
    the execution of tests.
    ]{{\includegraphics[width=0.45\columnwidth]{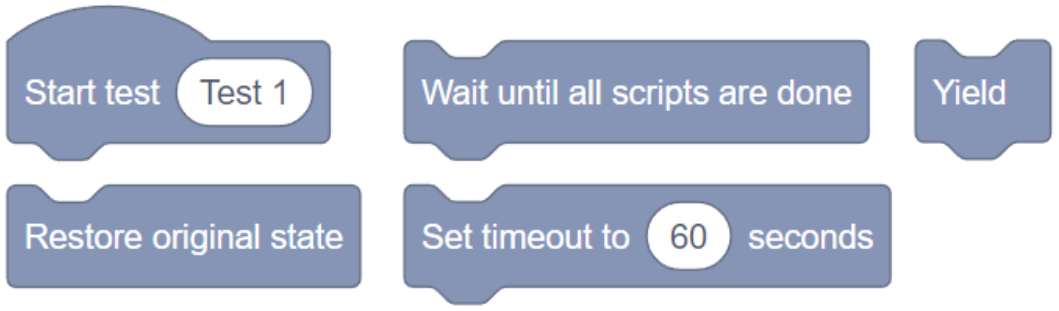}}}
    \hfill
    \subfloat[\label{fig:triggerBlocks}\centering Blocks for triggering event 
    handlers in programs.
    ]{{\includegraphics[width=0.45\columnwidth]{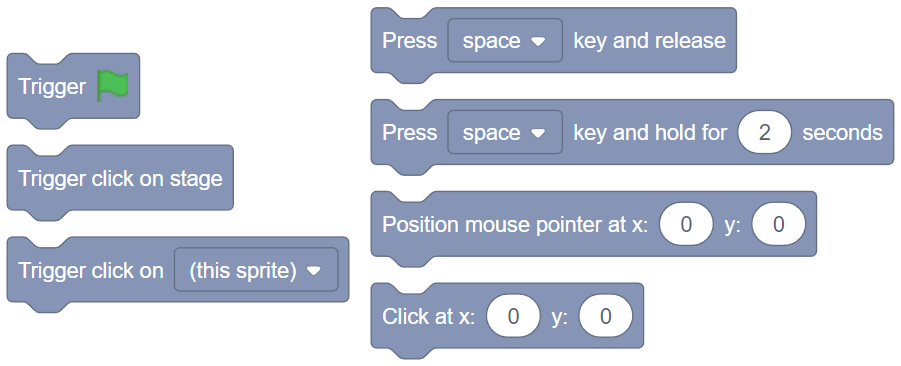}}}
    \\
    \vspace{1.2em}
    \subfloat[\label{fig:assertionBlocks}\centering Assertion blocks for 
    evaluating the program behaviour.
    ]{{\includegraphics[width=0.45\columnwidth]{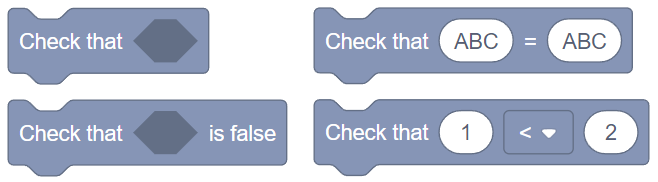}}}
    \hfill
    \subfloat[\label{fig:reporterBlocks}\centering Reporter blocks providing 
    access to sprite attributes.
    ]{{\includegraphics[width=0.45\columnwidth]{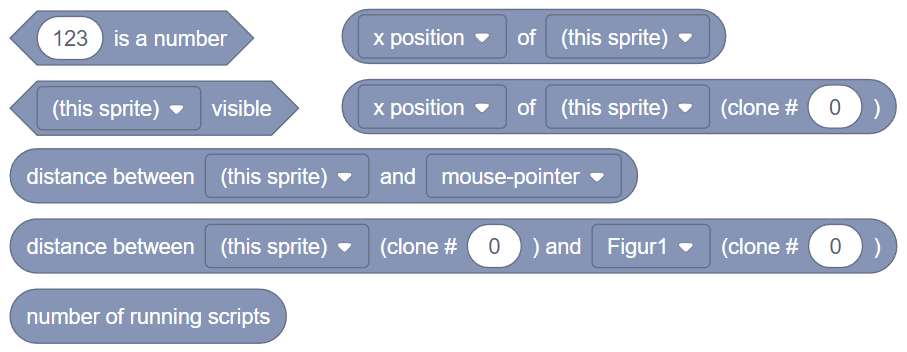}}}
    \caption{New blocks introduced for testing.}
    \label{fig:newBlocks}
    \vspace{-1.5em}
\end{figure}

\Cref{fig:controlBlocks} showcases the set of blocks added to manage the control flow of test scripts. The
\setscratchC{\begin{scratch}\testinit{Start test \ovalnum{Test 1}}\end{scratch}} block
incorporates a wide range of functionalities: First, it enables the
user to assign a specific name to the test, personalising the testing
process and improving usability in the presence of multiple test
scripts. Additionally, it safeguards against accidentally starting
multiple tests simultaneously by preventing the start of a new test
while another is currently active.  The block also implicitly
initiates a test timeout of five seconds to ensure that tests always
end within a reasonable time and do not get stuck in infinite
loops. Furthermore, the start block temporarily disables real mouse and
keyboard input to prevent intentional or accidental user interference
during test execution. Finally, the block is responsible for creating
a snapshot of the project's current state, which can be restored after
the test execution. This snapshot contains both the program state
(e.g., sprite attributes and variable values) and execution state
(running scripts and their execution progression).


At the end of each test script,
the \setscratchC{\begin{scratch}\testreset\end{scratch}} block may be inserted to
reset the project to its initial state, leveraging the snapshot taken
at the start of the test execution. The restore block ensures that
subsequent tests are not affected by state changes caused by
previously executed tests, which may otherwise cause flaky testing
behaviour~\cite{gruber2021empirical}. Placing this block at the end of
a test script is not mandatory; if omitted, the initially created
snapshot is discarded and any modifications to the project state are
kept, mirroring the behaviour of regular scripts that change the
program state.

The \setscratchC{\begin{scratch}\blocktest{Set timeout to \ovalnum{60}
    seconds}\end{scratch}} block allows users to set or refresh the
timeout that is started at the beginning of each test script. By
default, the timeout is set to five seconds, which is both
sufficiently long for most test scenarios and short enough for a swift
execution of entire test suites.  Explicitly setting the timeout is
particularly valuable for tests that are expected to exceed that
default timeout duration.  Users can place this block at any point
within the test script, enabling them to refresh the timeout during
the execution of the test and assign timeouts
to specific parts of their testing scripts.

\begin{figure}[!tbp]
    \centering
    \includegraphics[width=\columnwidth]{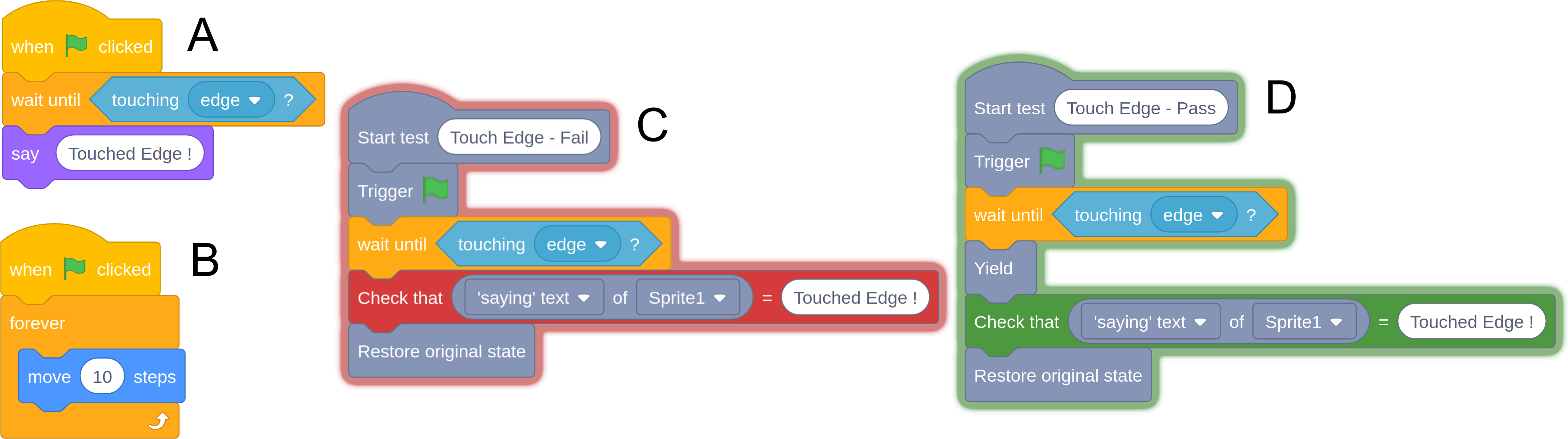}
    \caption{The passing test D uses a \emph{Yield} block to give execution precedence to script A.}
    \label{fig:yield}
\end{figure}

The \setscratchC{\begin{scratch}\blocktest{Yield}\end{scratch}} block addresses the
challenges of parallel script execution in the \Scratch \VM. As
explained in \cref{sec:background}, the \VM imitates parallel processing
by executing all active script threads $T$ in sequence, with each
active script $t_a$ being executed until a yielding block is
encountered or the end of the script is reached. This approach to
parallelisation may however require test scripts to stop their
execution temporarily in order to give precedence over the execution
to other active scripts. \Cref{fig:yield} illustrates this challenge
with an example: consider the two tests that are supposed to assess
whether a sprite displays the text \emph{Touched Edge!} after touching
the edge of the \Scratch stage.  The test script \emph{C} triggers the
green flag event and defers further execution of the test until
\setscratchA{\boolsensing{touching \ovalsensing*{edge} ?}} evaluates to true.  Since
the green flag event is triggered, program script \emph{B} starts to
continuously move the sprite towards its heading direction, eventually
causing the sprite to touch the edge of the stage after a few loop
iterations. This causes a race condition: If test script \emph{C}
continues \emph{before} the program script \emph{A}, it immediately
evaluates the assertion to check whether the sprite is displaying
text, which would fail because \emph{A} has not yet executed the block
to display the text. However, since the execution order of scripts is
unpredictable, the race condition could also have resulted in program
script \emph{A} taking precedence over \emph{C}, resulting in a passed
test. To avoid such ambiguous testing behaviour,
the \setscratchC{\begin{scratch}\blocktest{Yield}\end{scratch}} block can be used to
explicitly pause the execution of the test script, yielding to all
other active scripts before resuming the test. Thus, inserting this
deferring block after the \setscratchC{\begin{scratch}\blockcontrol{wait until
    \boolsensing{touching \ovalsensing*{edge} ?}}\end{scratch}} in test
script \emph{D} ensures that the block responsible for displaying the
text is guaranteed to be executed before the test script checks
whether the sprite displays the text.
\begin{figure}[t]
    \centering
    \includegraphics[width=.75\columnwidth]{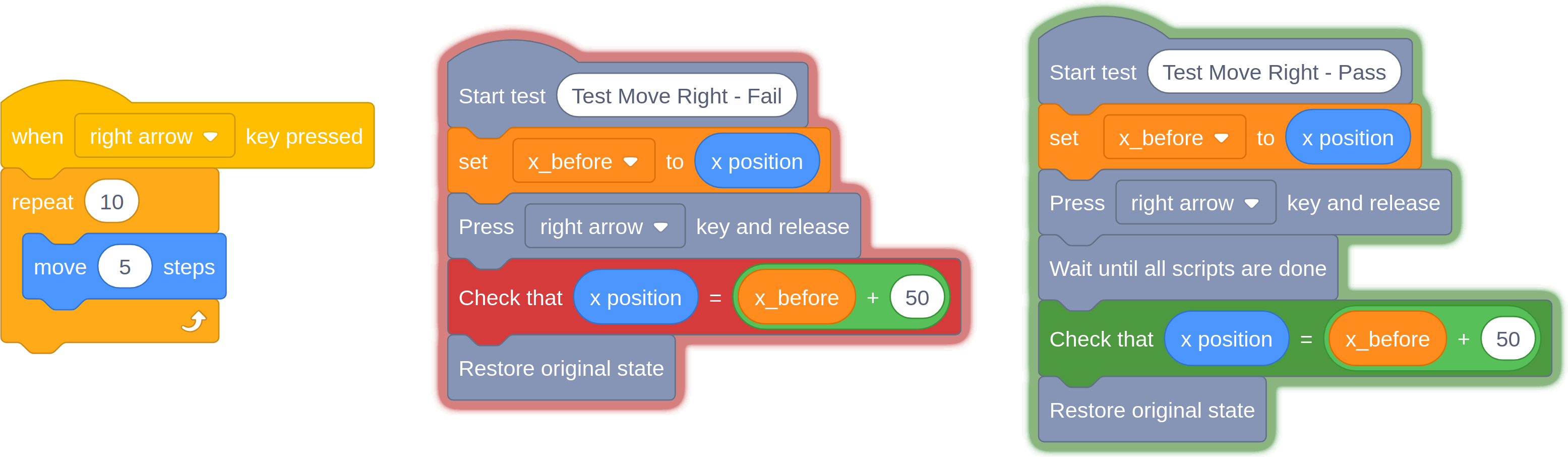}
    \caption{The passing test waits until the script containing the movement logic has fully completed.}
    \label{fig:waitScripts}
\end{figure}

In some scenarios, yielding the execution only once may not be enough. Consider the
two tests in~\cref{fig:waitScripts} that are responsible for checking whether
the corresponding sprite moves to the right when the user presses the right
arrow key. The left test fails because, even though triggering the script containing the movement instruction (and implicitly yielding to it), the sprite has only moved five steps to the right when the (failing) assertion block is evaluated. The sprite has not yet moved 50
steps, as the repeat block
also entails an implicit yield at the end of its body,
which transfers the execution priority back to the test script after its enclosed move block
has been executed once. Avoiding the need to insert nine
\setscratchC{\begin{scratch}\blocktest{Yield}\end{scratch}} blocks, we added the
\setscratchC{\begin{scratch}\blocktest{Wait until all scripts are done}\end{scratch}} block,
which defers the execution of a script until all other program
scripts have stopped. Hence, we can fix the test by inserting this control
block before the assertion to ensure
the loop in the program script has been repeated to completion, causing the sprite to
move 50 steps.

\subsection{Trigger Blocks}
\label{sec:trigger}

As \Scratch is highly event-driven, testing \Scratch programs requires
blocks dedicated to triggering the respective event handlers (shown
in~\cref{fig:triggerBlocks}) by sending corresponding events to the
\VM.  Program scripts in \Scratch usually start with a hat block that
listens to specific events (such as
\setscratchB{\begin{scratch}\blockinit{when this sprite clicked}\end{scratch}} ) and invokes successive blocks in that
script upon receiving the respective event.  To initiate these scripts programmatically, the trigger
blocks of our extension facilitate sending user input and events to
the \Scratch \VM.  Intuitively, users will expect the execution
priority to switch to newly invoked scripts after a trigger block is
executed. Due to the scheduling logic of the \VM
(cf.~\cref{sec:background}) however, the current script would continue
running.
To ensure that the activated script is executed \emph{before} the test
script proceeds with evaluating the program state, every
trigger block would have to be succeeded by
a \setscratchC{\begin{scratch}\blocktest{Yield}\end{scratch}}
block~(\cref{sec:control}).  To avoid bloat in test scripts, each
trigger block implicitly includes a yielding instruction, which
guarantees that activated scripts are executed before the test
execution is resumed and users do not have to clutter test scripts
with yield blocks.

\subsection{Assertion Blocks}
\label{sec:assertion}

The assertion blocks showcased in~\cref{fig:assertionBlocks} provide the major
purpose of every test, checking whether the program behaves as expected.
Whenever those assertion blocks evaluate a condition, their colour changes to
reflect the outcome: green signals that the condition has been fulfilled,
while red implies that the program has failed to meet the specified condition,
indicating a failure. This colour-coding system makes it straightforward for
users to ascertain the reason for the success or failure of their test scripts
at a glance. We consider a test passed if all its assertion blocks are
satisfied and deem a test failed if any of them have a negative outcome.

In principle, all assertions could be implemented using the
\setscratchC{\begin{scratch}\blocktest{Check that \testoperator{ }}\end{scratch}} block in combination with native \Scratch blocks of the \emph{operator}
category. However, the additional blocks improve the ease of use and
readability of tests since they embody some frequently used
assertions, including comparisons of equality and numerical
values. Furthermore, specialised assertions allow us to provide
a more detailed explanation of why a test case failed.

\subsection{Reporter Blocks}
\label{sec:reporter}

By default, the \Scratch programming environment provides access to
most sprite attributes, with different properties being spread across
multiple blocks. For instance, the native \setscratchA{\ovalmove{x position}} block
enables users to access the horizontal position of the current
sprite, but there are no default blocks that allow users to check if a
sprite is visible or is displaying some text. However, in a testing
scenario, users require fast and easy access to a multitude of
relevant sprite attributes, not only with regards to the currently
active, but \emph{all} sprites. In order to improve the testing
experience, we added the diverse set of reporter blocks shown
in~\cref{fig:reporterBlocks} to our testing framework. These
blocks allow users to easily access a sprite's position, direction, costume
number, costume name, size, volume, displayed saying/thinking text,
sprite clones, and the number of currently running scripts. In
addition, these blocks enable users to select a reference to the
current sprite instead of a specific name, which simplifies the reuse
of test scripts across different sprites. All reporter blocks are
designed to be compatible with existing \Scratch blocks and can also be used in
regular program scripts.


\section{Creating and Executing Tests}
\label{sec:CreatingAndExecutingTests}

\subsection{Creating Tests}
\label{sec:creatingTests}

From the user's perspective, block-based tests can be created
similarly to regular scripts by either utilising the \emph{drag-and-drop}
functionality from the toolbox, duplicating existing tests on the
workspace or copying tests across different sprites. Leveraging the
extension support of the \VM, these test blocks are designed to
integrate with the existing scope of functionality. Therefore, test
blocks can be combined with regular \Scratch blocks to offer a
powerful toolset for its users. For instance, in order to check some
condition within a test case, conditional blocks that are part of the
\Scratch \emph{operator} category can be inserted into an assertion
block that seamlessly evaluates the native condition.  Just like in
regular scripts, native blocks within test scripts can implicitly
refer to the currently selected sprite. The reporter blocks introduced
in this extension~(\cref{sec:reporter}) however have an option to
explicitly select the concerning sprite.

\begin{figure}[!tbp]
	\centering
	\includegraphics[width=0.65\columnwidth]{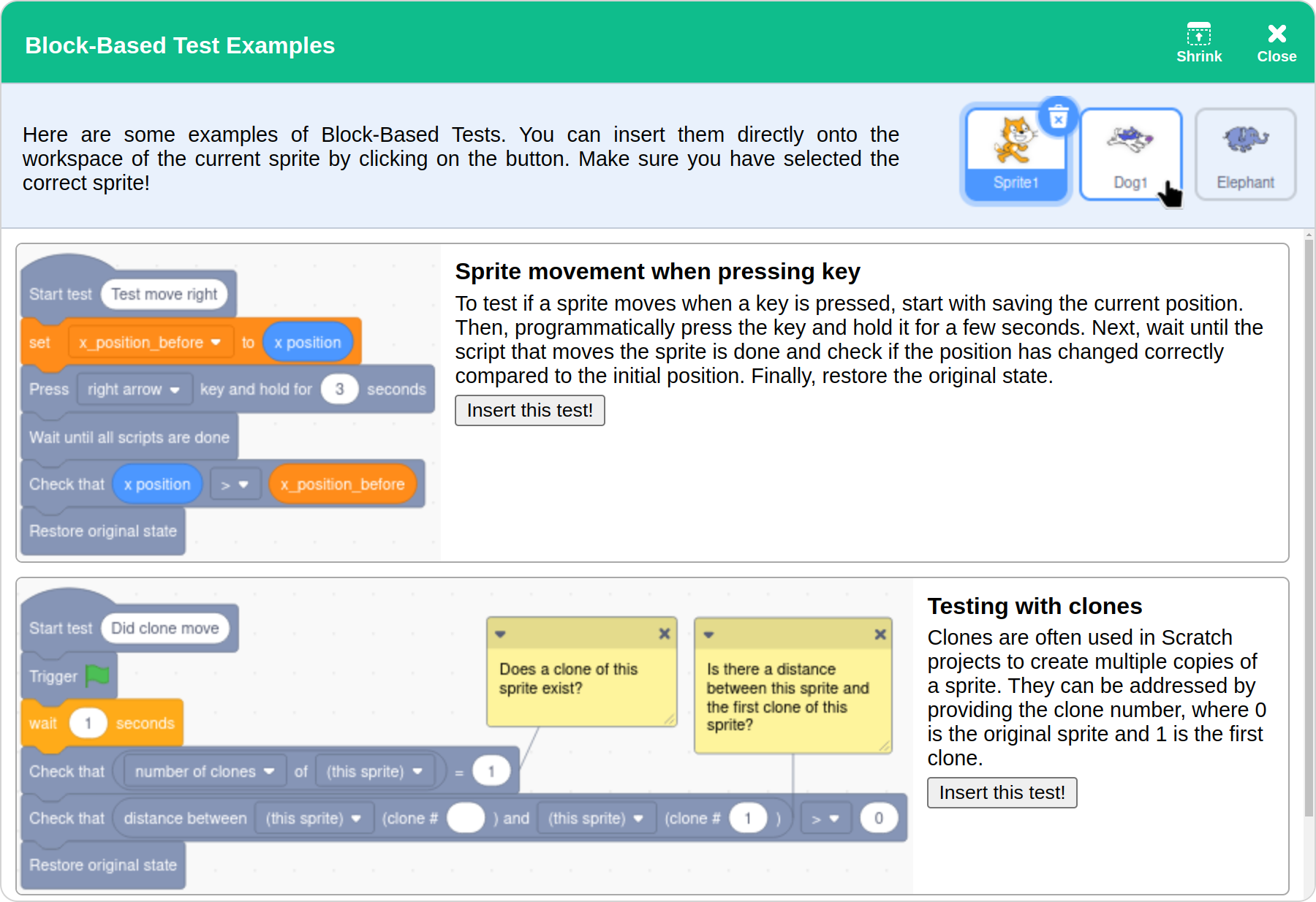}
	\caption{The examples window showcases tests for common scenarios and
		allows users to import them into their project.}
	\label{fig:exampleWindow}
\end{figure}

Since writing tests may be challenging for programming beginners who have never
tested programs before, our tool includes a dedicated examples window (shown
in~\cref{fig:exampleWindow}) to ease the introduction to block-based testing.
This window grants users access to a curated selection of pre-constructed tests.
These examples span a variety of common scenarios, such as testing the movement
of sprites upon arrow key activation, and provide practical help for
users to understand and implement testing strategies. They can be
imported to the workspace by pressing a single button. This
feature significantly lowers the barrier to creating tests by eliminating the
need to develop them from the ground up, thereby making the first steps into
testing more approachable and less intimidating for users. Since the template
tests are inserted as editable scripts, users can customise them according to
their needs.

\subsection{Test Execution}
\label{sec:testExecution}

\begin{figure}[!tbp]
	\centering
	\includegraphics[width=0.5\columnwidth]{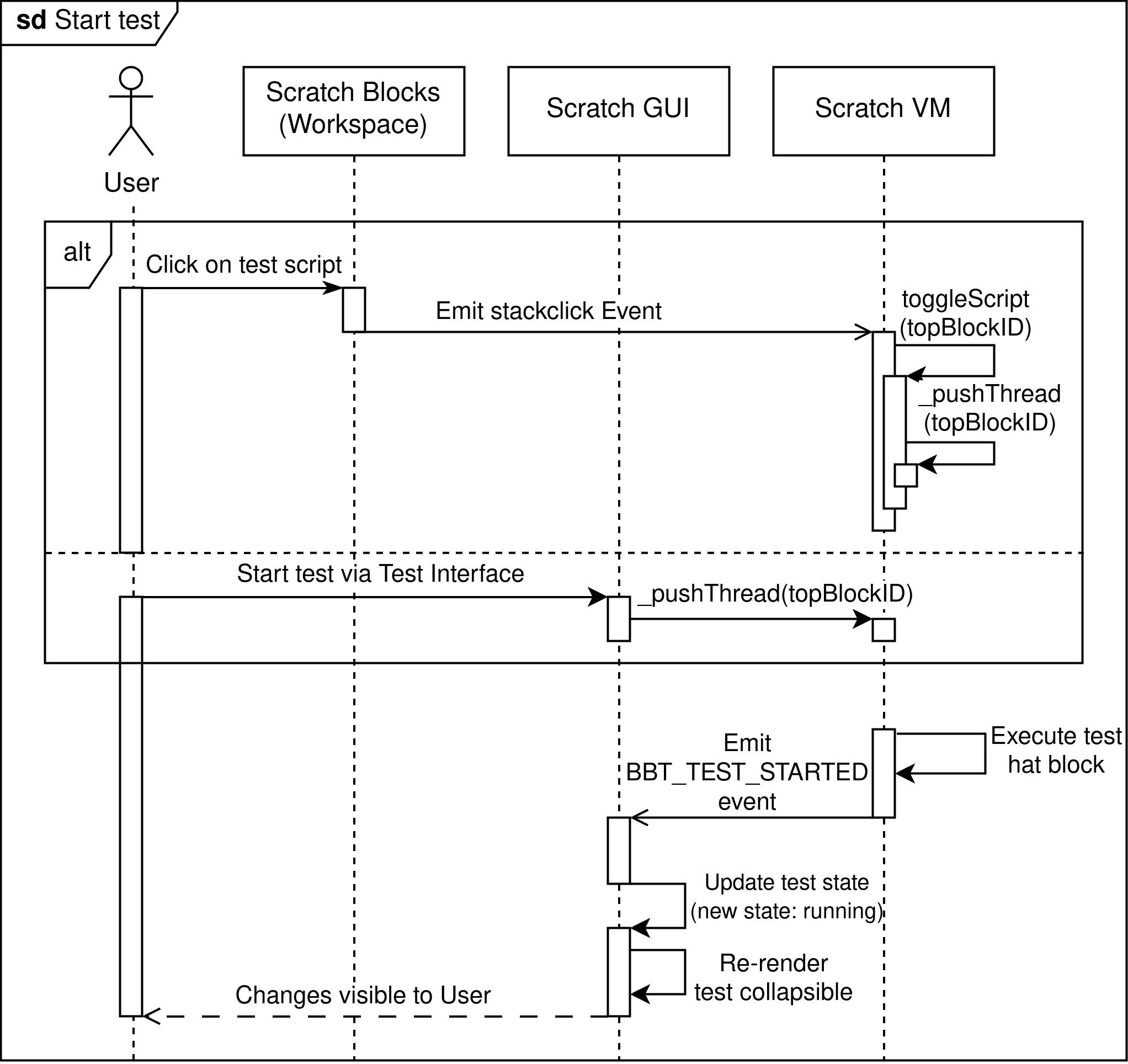}
	\caption{Simplified sequence diagram showing the interaction of \Scratch 
		components when a test is started.}
	\label{fig:startTest}
\end{figure}

A test script can be initiated by either directly clicking on it
in the workspace or by pressing the run button of an individual test (or the complete test suite) within the test 
interface~(\cref{sec:integration}). As described
in~\cref{fig:startTest}, the workspace issues a \emph{stackclick} event when a (test) script is clicked, to which
the \VM responds by executing the \emph{toggleScript} function. This function
then pushes a new thread, which manages the execution of the clicked test
script, into the list of active threads. In contrast, if a single test or the
entire suite is started through the test interface~(\cref{fig:testInterface}), the respective threads are
directly pushed to the list of active threads. As soon as the activated thread of the test is executed
by the sequencing function of the \VM, the \emph{BBT\_TEST\_STARTED} event is fired in
order to notify the user interface about the started test. The user interface
reacts to this event by updating the tests' status in the test
interface to \emph{running} and clearing any
errors from previous test executions.

In the ongoing execution process, test threads are treated like all other
program-related threads and do not have special precedence during the
execution, which might require the use of test-related control blocks in
specific scenarios~(\cref{sec:control}). Whenever it is the turn of a test
thread, the test is executed like any other program script by performing the
actions of the respective blocks that are hosted by the script. In case one of
the assertion blocks introduced in~\cref{sec:assertion} is executed, the test
interface is notified about the evaluation result, enabling it
to update the visualisation of the test in both the workspace and the interface itself with a green or red colour for a passed
or failed test, respectively. While tests are running, real user input and interactions with
testing functionalities are temporarily disabled to
avoid ambiguous testing behaviour.

A running test script can end due to one of three reasons. First, a
test may conclude naturally after the test script has been fully
executed, which causes the respective thread to be retired by the
\VM. Secondly, a test can come to an early end if the user decides to
abort the program execution by clicking on the
native~\raisebox{-3pt}{\inline{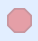}}button, as this causes the \VM to
retire every active thread immediately. Finally, a test may be
interrupted if it encounters a timeout that is set implicitly by the
\setscratchB{\begin{scratch}\testinit{Start test \ovalnum{Test 1}}\end{scratch}} block or
explicitly by the \setscratchC{\begin{scratch}\blocktest{Set timeout to
    \ovalnum{60} seconds}\end{scratch}} block. In order to ensure that
successive test cases of a test suite are not impacted by changes in
the program environment that may result from a partially executed test
that ran into a timeout, the program state is reset using the snapshot
of the \setscratchB{\begin{scratch}\testinit{Start test
    \ovalnum{Test1}}\end{scratch}} block~(\cref{sec:control}),
regardless of the presence of a \setscratchC{\begin{scratch}\testreset\end{scratch}}
block. Furthermore, all mouse and keyboard inputs are reset to prevent
lingering inputs from affecting successive test executions. Note that
failing assertion blocks do not lead to early termination and test
execution will continue regularly.

Following the completion of each test, irrespective of the manner of
its conclusion, the control buttons of the test interface are
re-enabled, and the processing of human user inputs resumes. If tests ended
naturally or due to timeouts, the test 
execution logic contained in the interface will automatically initiate
the next test in the sequence, provided the current test was started
as part of a test suite. However, for tests that were aborted,
advancing to the next test would counter the user's deliberate choice
to cease all activities by clicking the stop button and is therefore
avoided.


\section{Integrating Block-Based Tests into the \Scratch Environment}
\label{sec:integration}

\begin{figure}[!tbp]
    \begin{subfigure}[b]{.29\columnwidth}
      \centering
      \includegraphics[width=\columnwidth]{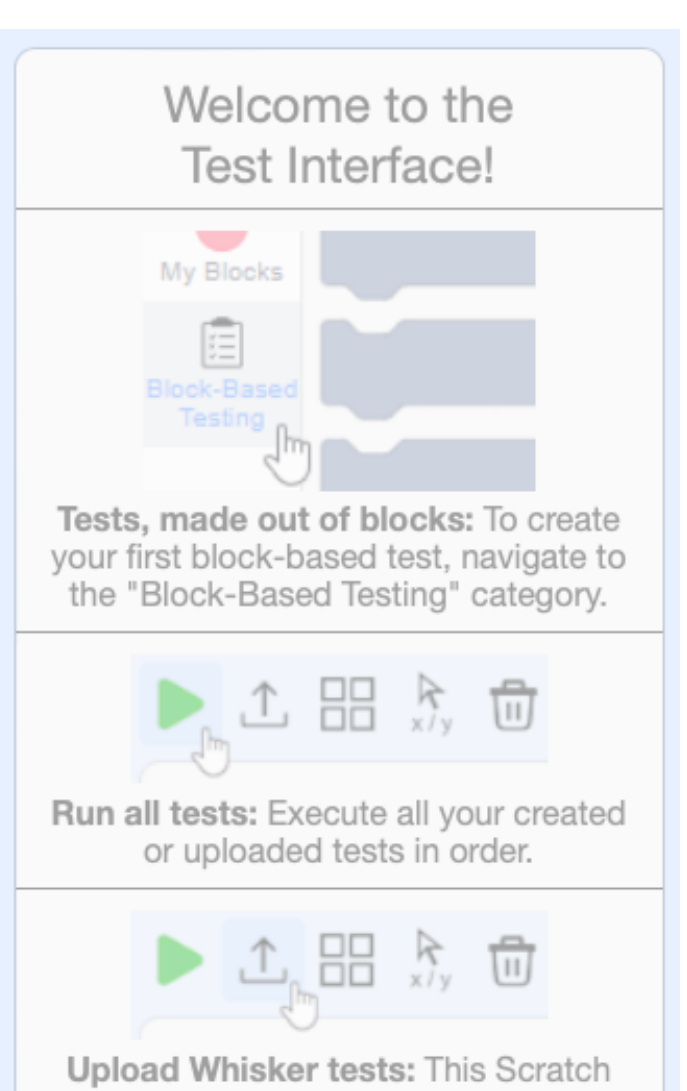}
      \caption{Introduction to block-based testing.}
      \label{fig:testIntro}
    \end{subfigure}%
    \hspace{1em}
    \begin{subfigure}[b]{.29\columnwidth}
        \centering
      \includegraphics[width=\columnwidth]{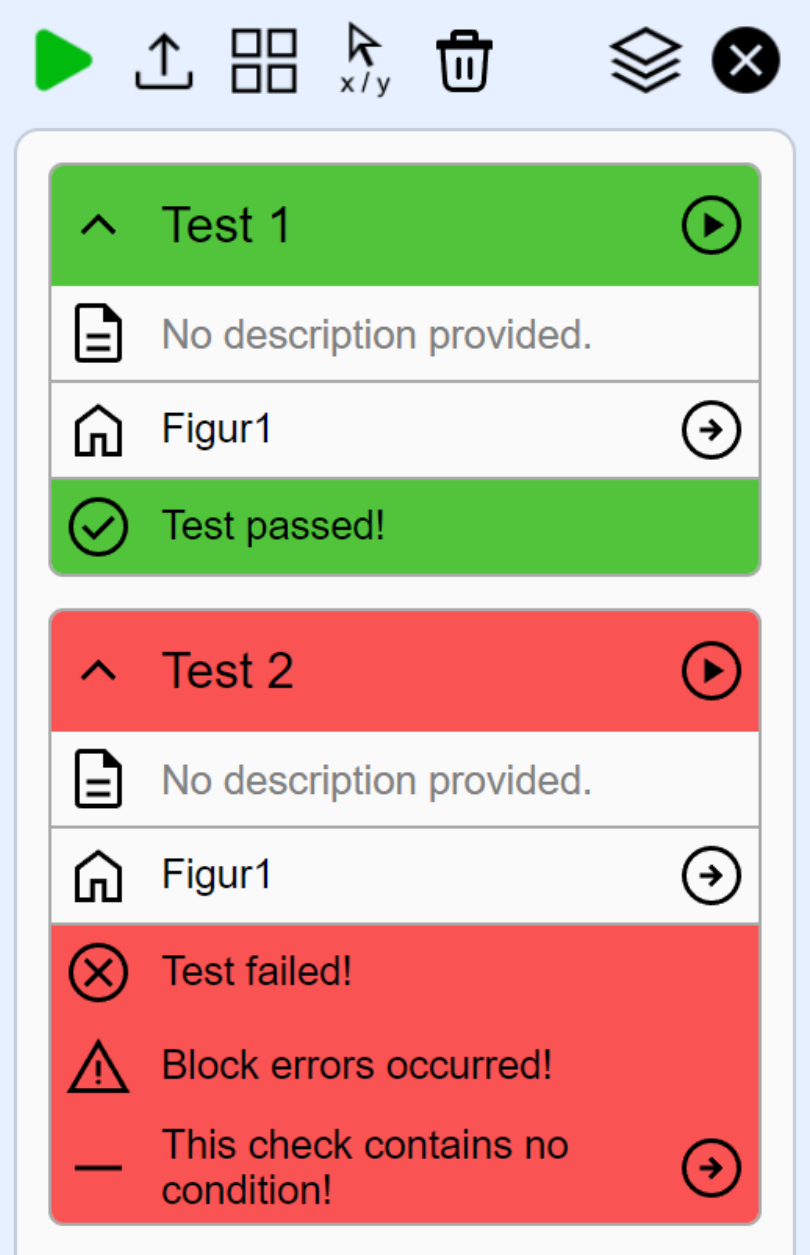}
      \caption{Overview of created tests.}
      \label{fig:testInterface}
    \end{subfigure}
    \caption{The test interface shows an introduction to block-based testing 
    if no tests exist, or an overview of all existing tests.}
    \label{fig:interface}
    \vspace{-1em}
\end{figure}

The test interface depicted in~\cref{fig:interface} is the main control component to manage and observe block-based tests. It is integrated within the 
existing components of the graphical user interface between the workspace and
the stage. A major focus was put on maintaining the editor's visual harmony by adhering to the design principles of
\Scratch to create an
interface that appears both native and intuitive.  

Initially, this area introduces users to the concept of block-based testing
through a quick start guide~(\cref{fig:testIntro}), supporting those new to test
creation. As soon as the first test is created, this section evolves to host
collapsible block entries for each test script~(\cref{fig:testInterface}). These
entries provide information about each test, including its name,
description, execution controls and contextual details such as the associated
sprite. Additionally, these entries display and highlight errors, using a
red/green colour scheme for immediate visual feedback on test outcomes. To
assist users in debugging, we combine the proposed block-based testing
framework with the interrogative 
debugger \Nuzzlebug~(\cref{sec:nuzzlebugIntegration}). 

The control bar, located at the top of the test interface (see ~\cref{fig:testInterface}), offers a range
of functionalities that cater to different testing needs, including running all
tests sequentially~\raisebox{-3pt}{\inline{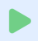}}(\cref{sec:testExecution}) and
displaying the examples window~\raisebox{-3pt}{\inline{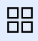}}introduced in
\cref{sec:creatingTests}. Using the~\raisebox{-3pt}{\inline{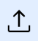}}button, users can upload
\Whisker tests~(\cref{sec:background}), which are then integrated as separate test cases to the test
interface like other block-based tests. Furthermore, we help users in writing tests that
involve precise coordinates by offering a tool~\raisebox{-3pt}{\inline{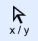}}that
dynamically displays the current coordinates of the mouse pointer within the
stage. The~\raisebox{-3pt}{\inline{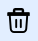}}button can be used to clear all test results
and the~\raisebox{-3pt}{\inline{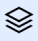}}button gives users access to execute created
tests on a batch of \Scratch programs~(\cref{sec:batchProcessing}). Finally,
the~\raisebox{-3pt}{\inline{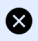}}button collapses the test interface area to avoid
cluttering the \Scratch interface if users are not interested in writing tests.

\subsection{Debugging Test Outcomes}
\label{sec:nuzzlebugIntegration}

\begin{figure}[!tbp]
  \begin{subfigure}[b]{.49\columnwidth}
    \centering
    \includegraphics[width=\columnwidth]{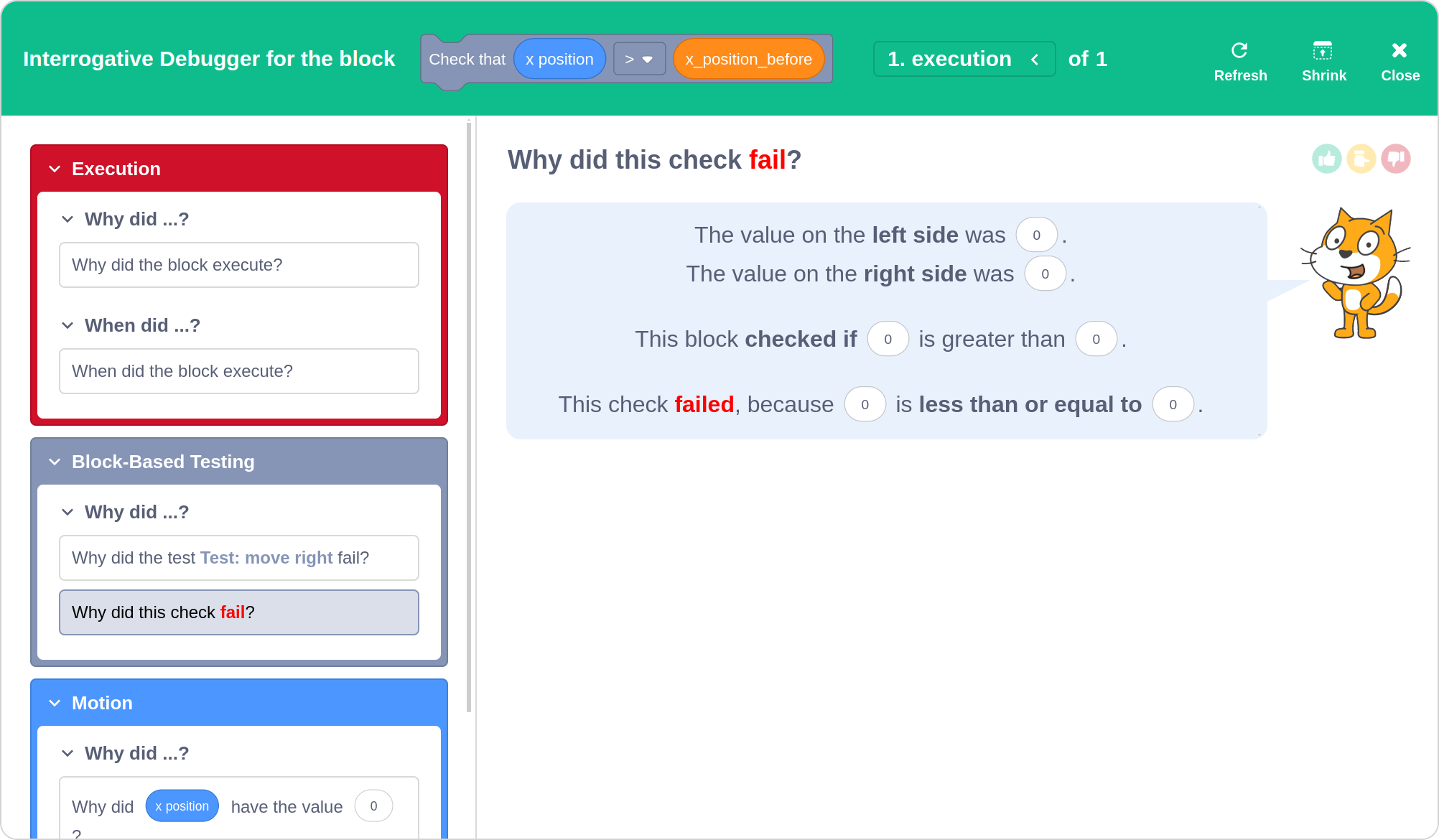}
    \caption{The interrogative debugger \Nuzzlebug is extended with a category to ask questions about the execution of tests.}
    \label{fig:nuzzlebugIntegration}
  \end{subfigure}%
  \hspace{.1em}
  \begin{subfigure}[b]{.49\columnwidth}
      \centering
    \includegraphics[width=\columnwidth]{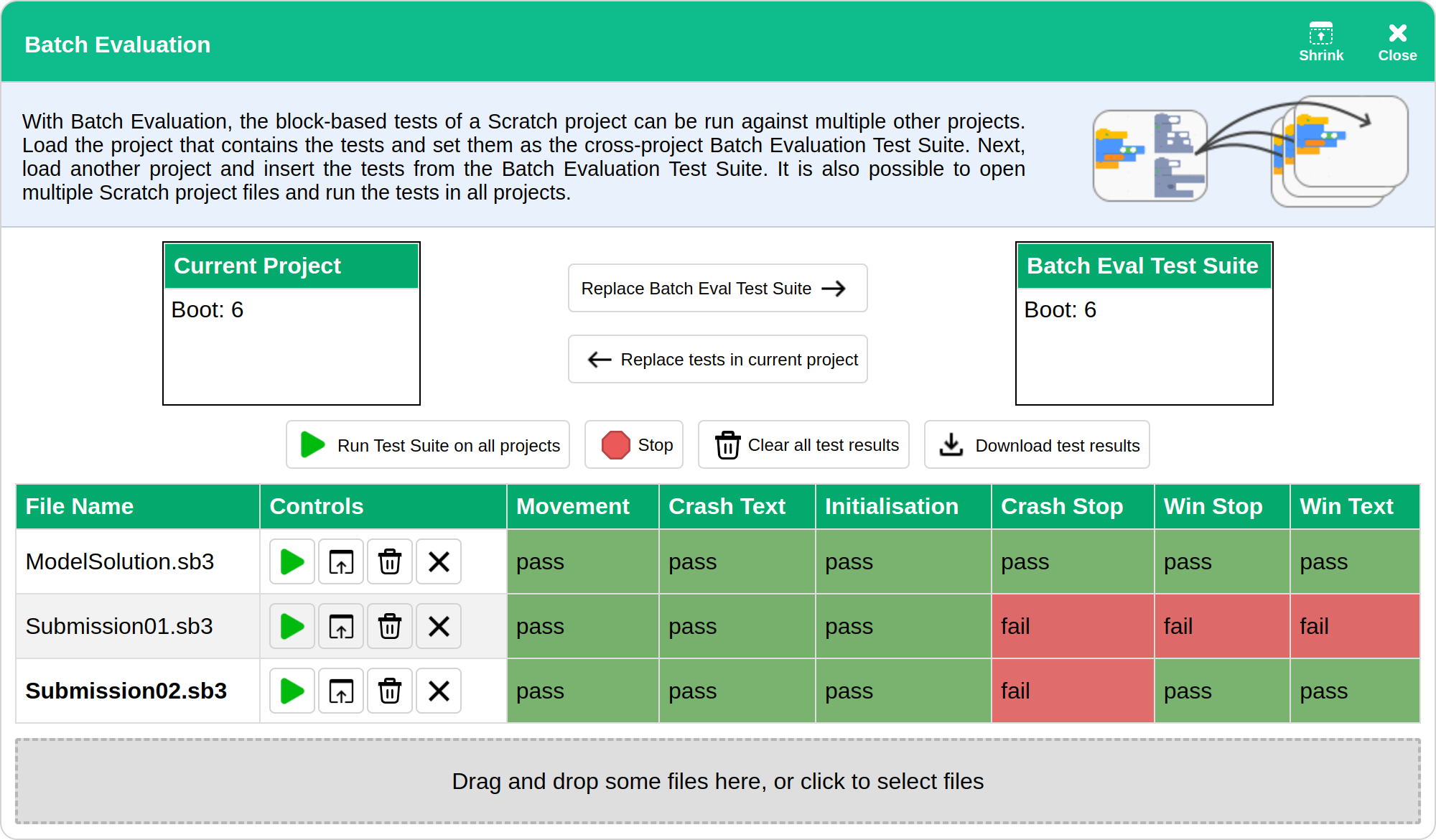}
    \caption{The batch evaluation window allows users to execute a test suite on multiple similar \Scratch programs.}
    \label{fig:batch}
  \end{subfigure}
  \caption{Extensions integrating block-based testing into the \Scratch environment.}
  \label{fig:extensions}
\end{figure}

Test results are visually indicated within the workspace by a glowing
effect around the respective test script, as well as through entries
within the test interface. If a test fails, the interface provides
detailed information about the nature of the error, pinpointing the
exact location where the issue occurred.  Users can quickly navigate
to the problematic block by clicking
the~\raisebox{-3pt}{\inline{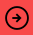}}button next to the error entry in the test
interface. To further assist users in debugging, we extended the
interrogative debugger \Nuzzlebug~\cite{deiner2024nuzzlebug} with
reasoning about block-based tests, as shown
in~\cref{fig:nuzzlebugIntegration}. This integration allows users to
pose questions about the reasons for failing tests and assertions. For
instance, \cref{fig:nuzzlebugIntegration} involves an example where
the debugger is queried about a failed numerical assertion block. The
interface outlines the necessary conditions for the assertion block to
succeed, thus offering a deeper insight and enabling users to fix
their implementation.

\subsection{Batch Testing of Scratch Programs}
\label{sec:batchProcessing}

By clicking on the~\raisebox{-3pt}{\inline{images/iconBatch}}button, users can launch the batch
processing window shown in~\cref{fig:batch},
which allows them to execute a block-based test suite across multiple \Scratch
programs. This functionality is orchestrated through the concept of a
cross-project \emph{batch test suite}, wherein tests from one project can be
designated as a suite to be applied across various others. Upon loading a
project into this environment, users can inject tests from the \emph{batch test
suite} into the new project while replacing all existing tests. The content of
the \emph{batch test suite} is not bound to the originating project file, but can
be dynamically updated or modified within any project into which it has been
previously injected, thereby enhancing the flexibility and adaptability of the
evaluation process and tailoring it to evolving testing requirements. 

The lower section of the batch evaluation window is occupied by the program
table, which presents a comprehensive listing of all \Scratch programs currently
loaded into the batch processing environment. New programs can be added to this
table via \emph{drag-and-drop} or by selecting \Scratch files using the
operating system's file browser. The second column of the table is equipped with
several controls, enabling users to automatically execute all tests from the
\emph{batch test suite} in a specific project, load projects into the editor,
remove file-specific test results or remove the complete file entry. Whenever
the batch evaluation procedure is initiated, each project of the file table is
first loaded into the \Scratch instance and then evaluated against the \emph{batch test suite}.
The result of each test case is then recorded in the file table, where each test
case is represented by a column named after the respective test case as defined by
the \setscratchB{\begin{scratch}\testinit{Start test \ovalnum{Test 1}}\end{scratch}} block.


\section{Evaluation}

To understand how block-based testing helps teachers assess students,
we consider the following research questions:
\begin{itemize}
\item \textbf{RQ1:} How accurately do the teachers' tests match the task description?
\item \textbf{RQ2:} How accurate is the teachers' final assessment?
\item \textbf{RQ3:} What do teachers think about block-based testing?
\end{itemize}

\subsection{Experimental Setup}
\label{sec:exp}

\subsubsection{Experiment Participants}

We conducted a pilot study with five in-training teachers without a
computer science background, introduced them to
\Scratch and refined the study setup based on
the insights. Data from the pilot study was not included in further
analyses.
The main study was conducted with 20 practising teachers
participating in professional training and 8 in-training teachers.
In the following, we will refer to all of them as teachers.

\subsubsection{Experiment Procedure}

Since only five teachers had used \Scratch before and the others had
little to no experience, we started with a two-hour introduction to
\Scratch, in which the teachers were tasked to implement smaller
exercises, a larger game, and a code understanding task explaining the
scripts of the \emph{Boat Race} tutorial\footnote{\url{https://projects.raspberrypi.org/en/projects/boat-race}}.
In a second session, we started with an introduction to block-based
testing, including an exercise to write a test for a task
from the introductory session. We included examples and exercises
using all test blocks and strategies needed throughout the study and
provided cheat sheets summarising the introduction.
Following the introduction, the teachers received the \emph{Boat Race}
\Scratch project and were tasked to write tests for
it. Since writing tests without knowing specific details about the
task would be unrealistic in a teaching scenario, we specified six
functionalities to be tested based on the
tutorial~\cite{obermueller2023tutorials}:
 \begin{enumerate}[label=(\alph*)]
 	\item The boat starts in the harbour with the costume ``normal''.
 	\item The boat follows the mouse pointer.
 	\item When the boat crashes into the wall, it says something.
 	\item After crashing into a wall, the boat's costume is ``damaged''.
 	\item When the boat reaches the beach, it says something.
 	\item After reaching the beach, the boat's costume is ``festive''.
\end{enumerate}
After all participants felt they had invested sufficient time in creating
their tests, we demonstrated batch testing~(\cref{sec:batchProcessing}) and explained how to
change and refine tests while inspecting student solutions. We then
supplied the 28 teachers with 21 student solutions for the \emph{Boat Race}
program~\cite{obermueller2023tutorials} and tasked them to assess the
six functionalities listed above for at least five solutions in the
remaining time of a total study duration of three hours. Although they were told to use their test results, they
were free to manually inspect the projects, change their test cases,
or simply provide assessment differing from the test
results. Assessment had to be submitted via an online form, where the
teachers could tick whether the student implemented each functionality
correctly or not, and add textual explanations to the given
\emph{correct/not correct} decisions. At the end of the experiment, the
teachers were given a survey consisting of nine questions, asking their
agreement on a 5-point Likert scale to questions regarding their
experience with the framework, and three free-text questions about issues
encountered, suggestions for improvement, and general remarks.

\subsubsection{Analysis Procedure}

\paragraph{RQ1:}
To determine the accuracy of the teachers' tests, we created a \emph{golden test suite} consisting of 6 tests (one for each required
functionality), which we manually refined to provide the correct
outcomes for all student solutions. We then executed each teacher's
test suite on the 21 student solutions, and calculated accuracy as the
matching ratio of pass/fail verdicts between our golden test suite and
the teachers' tests to the overall number of tested functionalities
(21 solutions $\times$ 6 functionalities). If a teacher did not write
a test for a functionality, we count this as a mismatch for each
student solution. We answer RQ1 by reporting the distribution of
accuracy values across all teachers.

\paragraph{RQ2:}
To determine the accuracy of the teachers' assessment provided via the
online form for each participant, we ascertain the ratio of
\emph{correct/incorrect} assessment results matching the
\emph{pass/fail} results of our \emph{golden} tests on the subset of
student solutions evaluated. We additionally consider the ratio of
\emph{correct/incorrect} assessment results matching the
\emph{pass/fail} results of \emph{their own tests} to determine
disagreement between test and assessment results.

\paragraph{RQ3:} 
In order to analyse the opinions of teachers on the block-based
testing framework, we summarise the 5-point Likert scale answers and
the free-text answers for each question of the exit survey.

\subsubsection{Threats to Validity}

Threats to \emph{internal validity} may result from using a 
predefined task, as teachers tend to perform better when working on a task
they created and understand well.
Threats to \emph{external validity} may stem from the number of
participants and the use of only one task and data set, limiting
how well results may generalise to other tasks and student
solutions. To counter this threat, we support replication by
providing all source code and materials online.
Threats to \emph{construct validity} can arise from our focus on
evaluation correctness, as we do not analyse the time needed or effects on other
assessment attributes, such as helpfulness and verbosity.



\subsection{RQ1: How accurately do the teachers' tests match the task description?}
\label{sec:results}

\begin{figure}[t]
	\centering
	\subfloat[\label{fig:acc_test_v_sample}\centering Test accuracy
     ]{{\includegraphics[width=.12\linewidth]{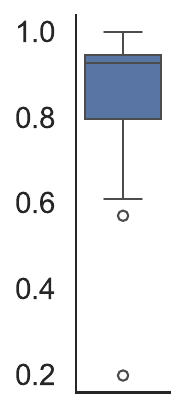}}}
		\subfloat[\label{fig:acc_test_v_feedback}\centering Assessment 
	agreement
	]{{\includegraphics[width=.12\linewidth]{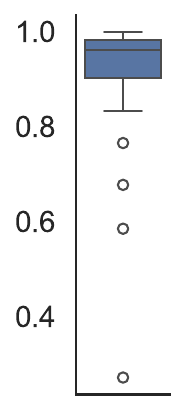}}}
	\subfloat[\label{fig:acc_feedback_v_sample}\centering
        Feedback 
        accuracy]{{\includegraphics[width=.12\linewidth]{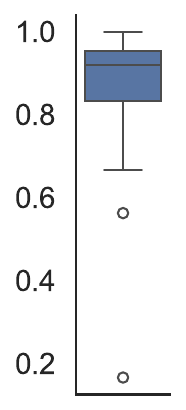}}}
        \subfloat[\label{fig:number_tests}\centering \#{}Tests
        ]{{\includegraphics[width=.12\linewidth]{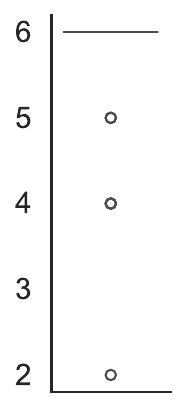}}}
	\subfloat[\label{fig:number_evaluated}\centering Solutions evaluated
	]{{\includegraphics[width=.12\linewidth]{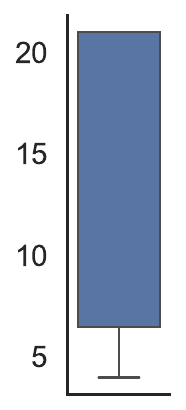}}}
    \subfloat[\label{fig:changes_worse}\centering Incorrect changes
    ]{{\includegraphics[width=.12\linewidth]{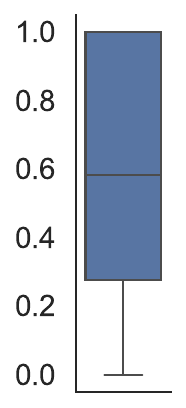}}}
	\caption{\label{fig:accuracies} The accuracies comparing teacher tests 
	and gold tests, teacher tests and assessment 
	and teacher assessment to the gold tests, the numbers of tests 
	written and students assessed, and the share of changes to the test results 
	not fitting the student.}
	\vspace{-1em}
\end{figure}

\Cref{fig:acc_test_v_sample} plots the distribution of accuracy values
for each of the 28 participants' tests compared to the golden test
suite. The median accuracy of \meanAccuracyTestSample confirms that
the tests cover the six specified functionalities very well most of
the time, and we conclude that teachers are able to write tests useful as a 
reliable measurement for automated assessment.

Although \cref{fig:number_tests} shows that most teachers wrote the 6
tests required to cover all aspects of functionality, there are a few
who did not (2 participants wrote only 5 tests, 3 wrote only 4, and
one participant wrote only 2). To
some degree, these missing tests contribute to the loss of overall accuracy in
\cref{fig:acc_test_v_sample} (median accuracy 0.94 when ignoring missing tests).
For four out of the six teachers who did
not write all tests, those tests they did write were good (overall accuracies between 0.61 and 0.80). Given
more time, they likely would have eventually produced the missing
tests.

The worst performing participant (accuracy 0.20), visible as a clear
outlier in \cref{fig:acc_test_v_sample}, only wrote tests for two
functionalities, and even these two tests were not very well suited for
assessing the targeted functionalities: One test checks whether the boat touches a brown wall immediately after the green flag is clicked (which should always fail) and the other test simply positions the mouse pointer without ever asserting anything (thus consistently passing).

The other outlier in \cref{fig:acc_test_v_sample} (accuracy 0.57)
wrote five (mostly fitting) tests, but one test, intended to
check whether the boat says something when colliding with the walls,
actually checks if the word \emph{apple} is longer than 50 characters, which is
nonsensical and always results in a failed test. Likely, this teacher
tried to check if the boat says anything but then simply forgot to
replace the default values of the \Scratch blocks.

\summary{RQ1}{Most teachers can write tests for common tasks after 
a short introduction to block-based testing.}

\subsection{RQ2: How accurate is the teachers' final assessment?}

While RQ1 confirmed the teachers' capability to write useful
tests, we now examine whether they also delivered proper assessments
in the end. \Cref{fig:acc_feedback_v_sample} indicates highly correct assessments with a median accuracy of
\meanAccuracyFeedbackSample. Thus, overall, the teachers provided correct
assessment.
\Cref{fig:acc_test_v_feedback} rates the agreement between the
teachers' assessment and the results of their tests at a median of
\meanAccuracyTestFeedback. This demonstrates that, by and large, the
teachers' final choices are in agreement with their tests. However, there
is a minimal decrease in the accuracy compared to the test results
(from 0.93 to 0.92), as the teachers sometimes disagreed with their
own tests.
All but four teachers made at least one change between test outcomes
and their assessment, resulting in \totalChanges cases where assessment
differed from the test outcomes overall. \Cref{fig:changes_worse}
illustrates that these changes slightly tend to make things worse
rather than better (median \meanWorse): \totalChangesWorse cases
changed correct test outcomes to incorrect assessment, and
\totalChangesBetter cases with incorrect test outcomes were
corrected.

Out of the \totalChangesWorse changes to a wrong assessment, a total
of \totalPassToFailWrong changes were passing tests for which 
some teachers apparently decided to be more strict than their block-based tests and our
golden test suite. For example, one teacher produced a test suite with
perfect accuracy, but then changed four test results as, in their eyes,
the student solution was not good enough and had room for improvement
(e.g., the boat gets teleported to the mouse pointer immediately after
the initialisation, so it is barely visible). Other examples of this
are two teachers who changed the assessment for the mouse following
functionality to \emph{not correct} even though the test passed,
because of a long waiting block before the boat starts its motion,
which probably was not the reaction the teachers expected. Another
teacher disagreed with the test result on the costume change to \textit{damaged}
when hitting the walls as \emph{correct}, as
the boat starts with the festive costume.

We also observed 82 cases where teachers overruled failing test
results and decided the outcome was acceptable. In
\totalFailToPassCorrect cases, this fixed an incorrect test outcome,
but in \totalFailToPassWrong cases, they accepted incorrect behaviour
as correct. While this may be influenced by misunderstandings, there are
probably also cases where teachers deliberately decided to be more
lenient.
Generally, in such cases of disagreement we would have liked teachers
to engage more with refining their test suites to match their actual
expectations, but it seems that in this aspect the
framework may need future improvements.

\Cref{fig:acc_test_v_feedback} shows the same outlier already
discussed for RQ1, who only wrote two tests: One of the tests always
\emph{passes} as it does not check anything, resulting in disagreement
with the assessment. Interestingly, no assessment was given on the
functionality tested by the other test, but the teacher instead
assessed a different aspect of functionality that was not tested at all.
Given that this teacher assessed the fewest (only four)
submissions (\cref{fig:number_evaluated}), the problem was likely not just 
struggles with our framework, but general lack of engagement with the
experiment or \Scratch.

The other outlier discussed for RQ1 can also be seen as the second to
lowest outlier in \cref{fig:acc_test_v_feedback}, mainly affected by
the test that always fails. This particular participant seemed to
focus on giving good student assessment by also evaluating the
functionality without tests. While this resulted in high accuracy of
the assessment (0.87, \cref{fig:acc_feedback_v_sample}), the agreement
between test results and assessment is comparatively low (0.59,
\cref{fig:acc_test_v_feedback}).

\begin{figure}[tb]
	\centering
	\subfloat[\label{fig:bad-control-flow}\centering Incorrect test.
	]{{\includegraphics[width=0.28\linewidth]{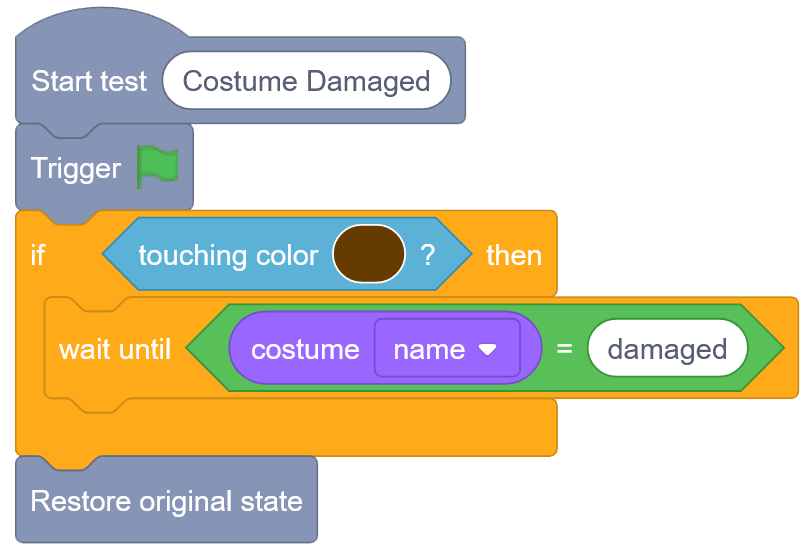}}}
	\hspace{0.01\linewidth}%
	\subfloat[\label{fig:bad-control-flow-fix}\centering Fixed test, yielding explicitly.
	]{{\includegraphics[width=0.24\linewidth]{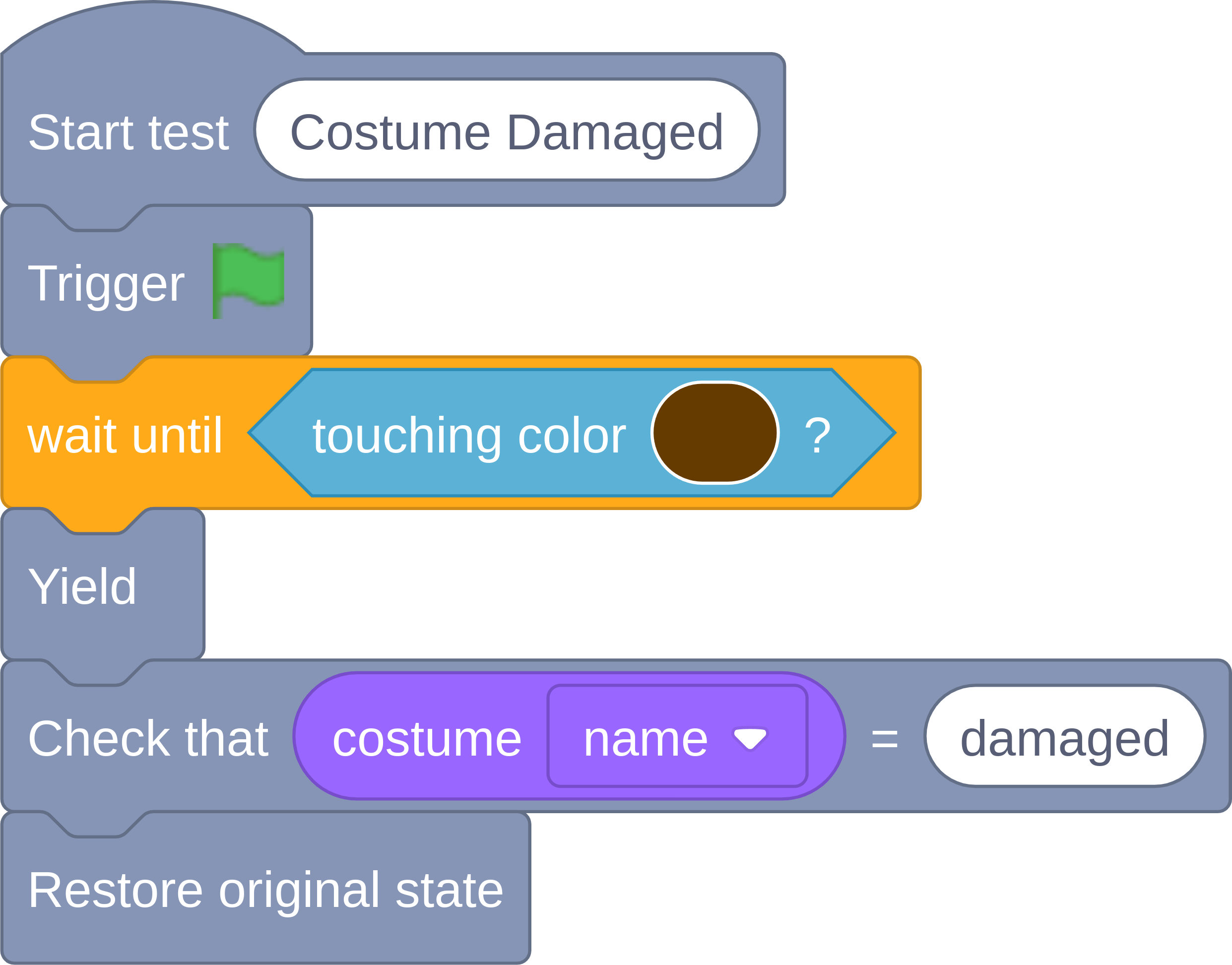}}}
	\hspace{0.01\linewidth}%
	\subfloat[\label{fig:bad-control-flow-fix2}\centering Alternative: implicitly yielding block.
	]{{\includegraphics[width=0.46\linewidth]{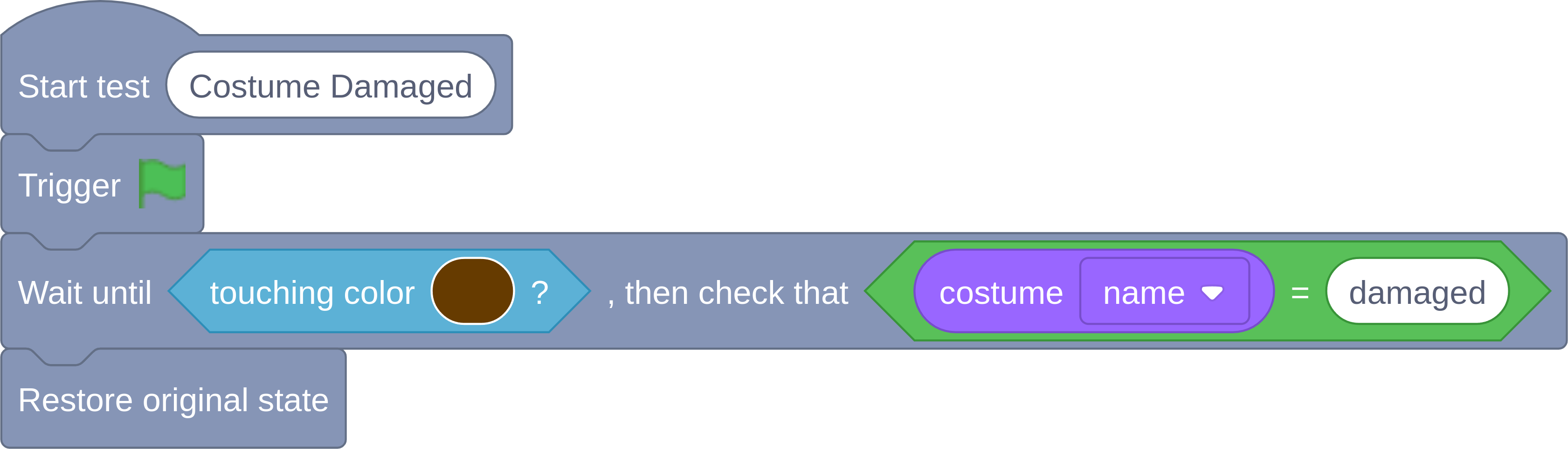}}}
	\caption{\label{fig:bad-control-flow-figure}
	The test attempts to check if the boat costume changes after 
	touching the colour of the wall; 
	if this does not occur within 5 seconds, the test will time out.
	However, this check is performed only \emph{once}
    after triggering the green flag,
    but the boat has not collided with the wall yet. Fixing the test includes waiting for the prerequisite condition and yielding afterwards.}
    \vspace{-0.5em}
		
\end{figure}

Another interesting outlier wrote six tests of which four
always fail, even for correct student solutions, resulting for RQ1
in a low accuracy of 0.68. This is caused by
tests performing a single collision check immediately after the green flag is activated (see \cref{fig:bad-control-flow-figure}), thus 
showing the common bug pattern \emph{Missing Loop Sensing}~\cite{fraedrich2020}. 
An assessment accuracy of
0.9 for this participant shows that they were able to come to mostly
correct conclusions despite their incorrect tests.

A further factor, contributing to the slightly lower accuracy of the
assessment compared to the tests, are 12 cases in which
teachers only provided an explanation for their assessment without
actually selecting \emph{correct/not correct}, with one even selecting
both options. In our analysis procedure, this is also counted as a
mismatch, as no definitive assessment was given. Such issues could be
overcome by extending the batch testing interface to a full assessment
tool, in which only test results deemed as incorrect need to be
changed by hand.

\summary{RQ2}{The teachers correctly assessed most functionalities,
  although they sometimes disagreed with their tests to be more strict
  or lenient without updating them.}

\subsection{RQ3: What do teachers think about block-based testing?}

\begin{figure*}[tb]%
	\centering 
	\includegraphics[width=0.7\linewidth]{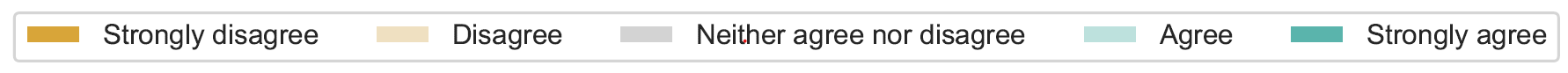}
	\subfloat[\label{fig:post-test1}\centering I am confident in using the
	block-based test 
	interface.]{{\includegraphics[width=.33\linewidth]{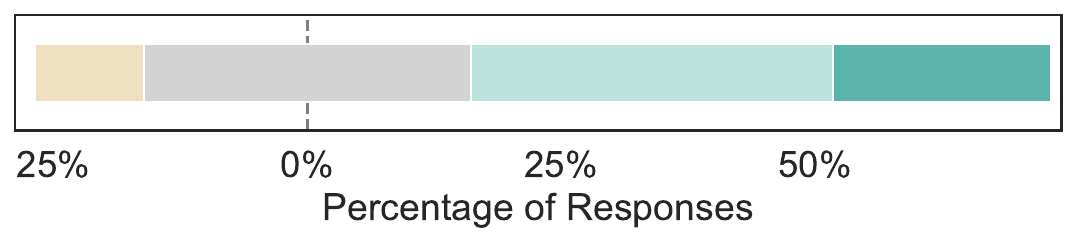}
	}}%
	\subfloat[\label{fig:post-test2}\centering I had problems using block-based 
	tests.]{{\includegraphics[width=.33\linewidth]{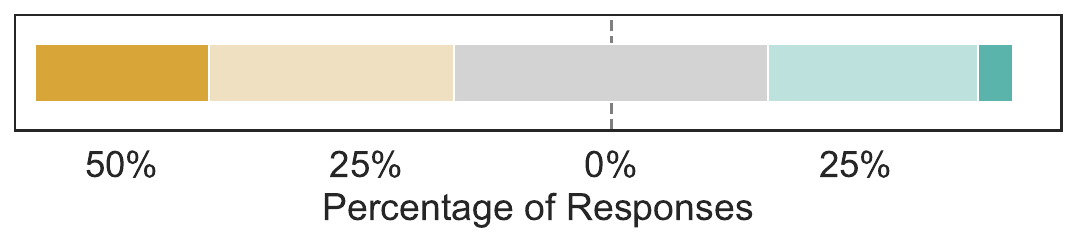} 
	}}
	\subfloat[\label{fig:post-test3}\centering I have not changed my tests once 
	they have been written. 
	]{{\includegraphics[width=.33\linewidth]{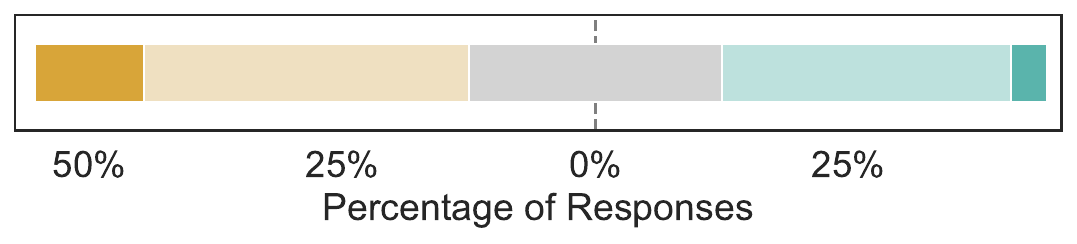} }}\\
	\subfloat[\label{fig:post-test4}\centering The more student solutions I 
	tested, the better my tests became. 
	]{{\includegraphics[width=.33\linewidth]{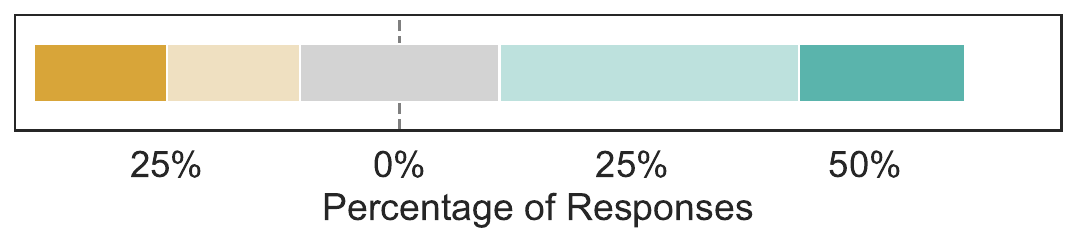} }}
	\subfloat[\label{fig:post-test5}\centering The block-based tests helped me 
	to find my way around the student project more quickly. 
	]{{\includegraphics[width=.33\linewidth]{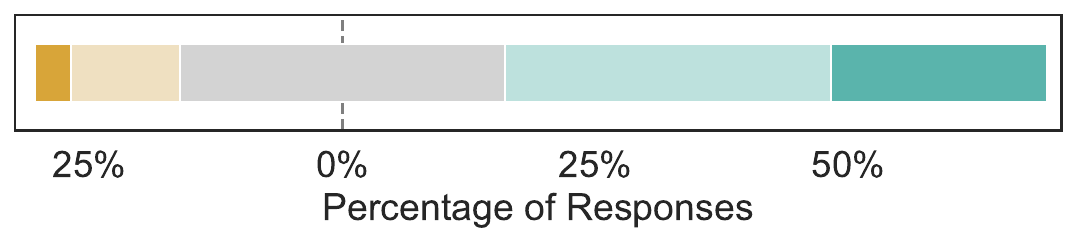} }}%
	\subfloat[\label{fig:post-test6}\centering Block-based tests help me to see 
	if the students have written program parts that work. 
	]{{\includegraphics[width=.33\linewidth]{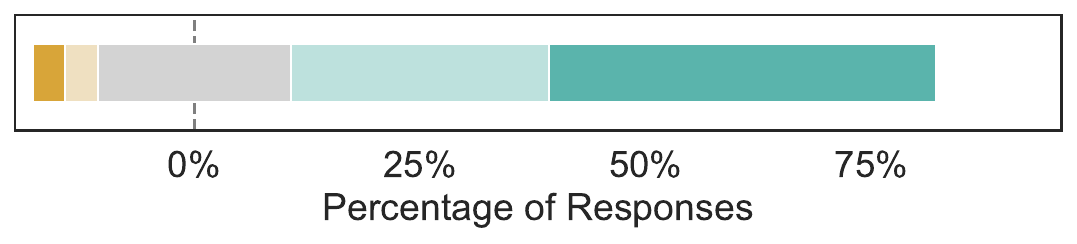} }}\\
	\subfloat[\label{fig:post-test7}\centering To make it easier to give 
	individual feedback, I would use block-based tests. 
	]{{\includegraphics[width=.33\linewidth]{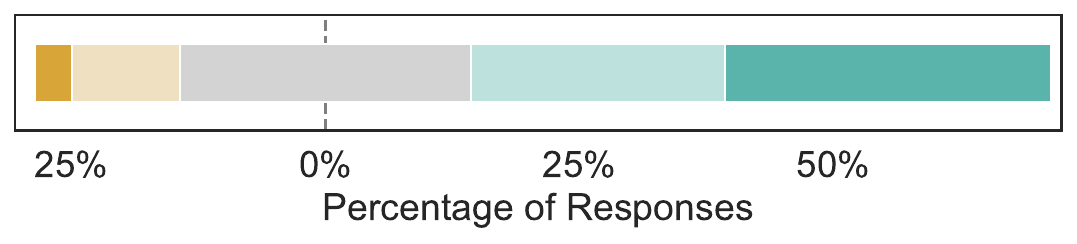} }}%
	\subfloat[\label{fig:post-test8}\centering I can use block-based tests that 
	others have written. 
	]{{\includegraphics[width=.33\linewidth]{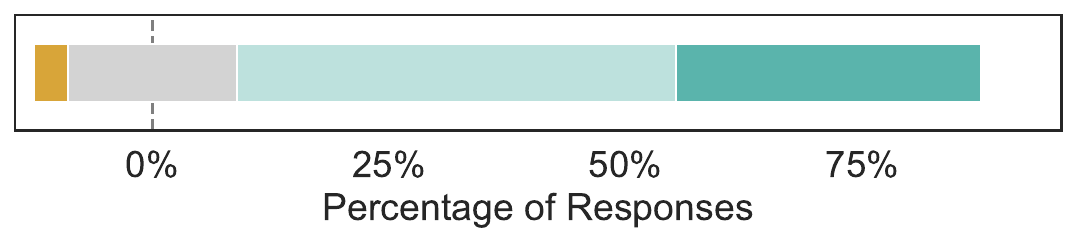} }}%
	\subfloat[\label{fig:post-test9}\centering I can write block-based tests 
	myself. ]{{\includegraphics[width=.33\linewidth]{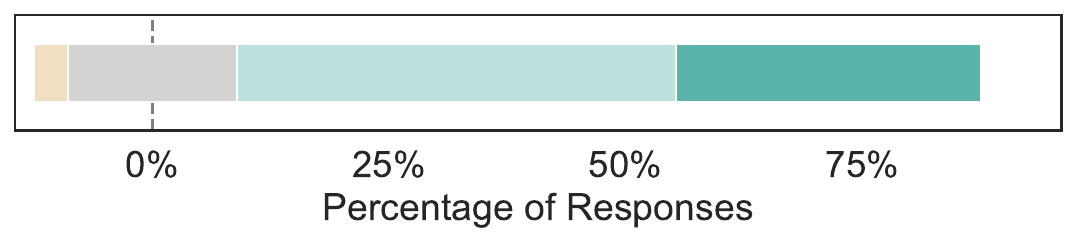}
	}}%
	\vspace{-0.8em}
	\caption{\label{fig:post-test}Survey results on how the teachers view 
		block-based tests.}
	\vspace{-1em}
\end{figure*}

\Cref{fig:post-test} summarises the responses to the survey questions on a
5-point Likert scale
and shows that the teachers are generally confident
in using the test interface itself, although \cref{fig:post-test2}
suggests that there were some problems with the concept of block-based
tests. The answers given for the free-text question \emph{If there
  were problems: What were they?}  mostly suggest that the teachers
wanted more time to familiarise themselves with the tool, and that
sometimes finding the right block was not as easy, which is a general
problem with the \Scratch interface. The teacher who had the lowest
accuracy for RQ1 and RQ2 stated that they faced problems not knowing
in which way they should test the functionalities. However, our
introduction and the cheat sheet contained all necessary information
about blocks and testing strategies. Maybe this is also an issue
rooted in the limited time. The one teacher who strongly agreed to
having problems with the tests had accuracies of 0.83 and higher in RQ1
and RQ2 and also strongly agreed to both being able to write tests
themselves and being confident in using the interface, so this
possibly was an incorrectly selected answer.

According to \cref{fig:post-test3}, there is a slight tendency to
change tests after the initial creation, and \cref{fig:post-test4}
also shows a slight tendency towards tests improving with more students
being evaluated. However, the opinions seem to strongly diverge on this
question.
Future work could investigate the use of approaches like mutation
analysis, simulating student mistakes~\cite{clegg2019simulating} in
order to guide teachers in producing adequate tests~\cite{clegg2020influence} 
prior to running them on student solutions.

While the majority of teachers consider block-based tests helpful to find
their way around the student projects (\cref{fig:post-test5}), a
single \emph{Strongly disagree} answer came from a teacher stating in
the additional comments that they sometimes were confused by the order
of their test cases in the interface and had a hard time looking up
test results because of that. This teacher also suggested that a
future version of the interface could include an option to reorder the
tests. This perception likely negatively influenced the reported
helpfulness of the tests. An even bigger majority
(cf. \cref{fig:post-test6}) agreed that block-based tests are
helpful to check if specific parts of student projects are
working. Most of the teachers also stated that they would use the
block-based tests to simplify providing individual feedback
(\cref{fig:post-test7}).  Again, the single \emph{Strongly disagree}
came from the teacher who had issues with the test order, which may
explain why they did not find it easier, as they had problems looking
up the test results.

Considering \cref{fig:post-test8} and \cref{fig:post-test9}, the vast
majority of teachers think that they can write block-based tests
themselves as well as use tests from other persons. However, in the
latter category one teacher selected \emph{Strongly disagree}, stating
that they never used tests from other persons as additional
comment. This may be a misunderstanding, as we intended to ask if they
would be willing to use it with other people's tests, even if they
would not be able to write tests themselves.

\summary{RQ3}{The block-based test framework was generally well received, but  
met suggestions for improvement.} 

\section{Conclusions}
\label{sec:conclusions}

Extending \Scratch with block-based tests supports
teachers when assessing student solutions and providing feedback. Furthermore, they
enable automated feedback and guidance, and an easy way to write tests
may be a valuable approach for introducing learners to the concept of
testing. In order to make this possible, we introduced a \Scratch
extension that provides test blocks and test controls,
integrated directly into the \Scratch user interface. Our initial
study provides evidence that teachers can work with tests to assess
students, and the teacher feedback is encouraging.

Having contributed to the foundations of testing in \Scratch, there are now
ample opportunities for future research:
\begin{itemize}
\item Our study revealed ideas for improving the interface itself, for
  example by adding a way to rearrange the displayed test order or to
  search for specific blocks in the \Scratch toolbox.
\item There is potential for adding further testing blocks. For
  example, inspired by Voeten's master thesis \cite{poke}, the ability
  to execute blocks as if they were part of another sprite's workspace
  would offer a practical approach to initialising a testing
  environment. Such blocks could eliminate the need to create specific
  scenarios through block-based, simulated user input (or manual
  changes) by allowing direct control over other sprites.
\item Recurring test patterns, such as verifying sprite movements,
  could be integrated as dedicated blocks of greater abstraction.  The
  current test-related blocks already allow low-level checking of any
  possible program behaviour, but higher-level blocks for specific
  issues could improve the efficiency of creating tests. To achieve
  this, it will be essential to gather data on tests created by
  teachers to identify common patterns.
\item While in this paper we focused on tests in which assertions
  check task-specific expected functionality, there may be general
  patterns of program misbehaviour or runtime errors that test
  executions can reveal, like sprites at the edge of the stage being
  unable to move further despite otherwise correct movement logic. Our
  testing framework provides the foundation for investigating such
  generalisable runtime checks to complement test assertions.
\item The example test in \cref{fig:bad-control-flow-figure} suggests
  that there can be common patterns of problems in block-based tests,
  akin to test smells~\cite{panichella2022test,van2001refactoring} in
  automated tests written in textual programming languages. Similar to
  how code smells can be defined and automatically detected for
  regular \Scratch code~\cite{fraedrich2020}, teachers and students
  may benefit from patterns and automated checks for block-based test
  smells.
\item While our initial motivation is to support teachers, there is
  also potential to integrate block-based tests directly into
  education, leading to the question of how learners should work with
  tests. For example, learners could be introduced to a test-driven
  development approach~\cite{edwards2003using} where they write tests
  prior to the code they intend to implement, or they could be taught
  to create automated tests to simplify debugging.
\item Research on explaining test failures to support debugging will
  be important, for example by providing more elaborate textual
  explanations, suggesting debugging hypotheses, or providing fix
  suggestions.
\item The test interface already provides an option to load and
  execute tests generated with the \Whisker automated test generation
  tool. It would be useful to extend automated test generation, as for
  example provided by \Whisker~\cite{deiner2023automated}, to directly
  create block-based tests rather than using other programming
  languages as output.
\end{itemize}

In order to support future research, we provide a full replication
package including all data and source code:
\urlstyle{sf}

\begin{center}
  \url{https://figshare.com/articles/dataset/25710288}
\end{center}

A live instance of our extended \Scratch version is available at:
\begin{center}
  \url{https://scratch.fim.uni-passau.de/block-based-testing}
\end{center}

\clearpage


\bibliographystyle{ACM-Reference-Format}
\bibliography{library}


\begin{thebibliography}{22}


\ifx \showCODEN    \undefined \def \showCODEN     #1{\unskip}     \fi
\ifx \showDOI      \undefined \def \showDOI       #1{#1}\fi
\ifx \showISBNx    \undefined \def \showISBNx     #1{\unskip}     \fi
\ifx \showISBNxiii \undefined \def \showISBNxiii  #1{\unskip}     \fi
\ifx \showISSN     \undefined \def \showISSN      #1{\unskip}     \fi
\ifx \showLCCN     \undefined \def \showLCCN      #1{\unskip}     \fi
\ifx \shownote     \undefined \def \shownote      #1{#1}          \fi
\ifx \showarticletitle \undefined \def \showarticletitle #1{#1}   \fi
\ifx \showURL      \undefined \def \showURL       {\relax}        \fi
\providecommand\bibfield[2]{#2}
\providecommand\bibinfo[2]{#2}
\providecommand\natexlab[1]{#1}
\providecommand\showeprint[2][]{arXiv:#2}

\bibitem[Bau et~al\mbox{.}(2017)]%
        {bau2017}
\bibfield{author}{\bibinfo{person}{David Bau}, \bibinfo{person}{Jeff Gray},
  \bibinfo{person}{Caitlin Kelleher}, \bibinfo{person}{Josh Sheldon}, {and}
  \bibinfo{person}{Franklyn Turbak}.} \bibinfo{year}{2017}\natexlab{}.
\newblock \showarticletitle{Learnable Programming: Blocks and Beyond}.
\newblock \bibinfo{journal}{\emph{Commun. ACM}} \bibinfo{volume}{60},
  \bibinfo{number}{6} (\bibinfo{date}{May} \bibinfo{year}{2017}),
  \bibinfo{pages}{72–80}.
\newblock
\showISSN{0001-0782}
\urldef\tempurl%
\url{https://doi.org/10.1145/3015455}
\showDOI{\tempurl}


\bibitem[Clegg et~al\mbox{.}(2019)]%
        {clegg2019simulating}
\bibfield{author}{\bibinfo{person}{Benjamin Clegg},
  \bibinfo{person}{Siobh{\'a}n North}, \bibinfo{person}{Phil McMinn}, {and}
  \bibinfo{person}{Gordon Fraser}.} \bibinfo{year}{2019}\natexlab{}.
\newblock \showarticletitle{Simulating student mistakes to evaluate the
  fairness of automated grading}. In \bibinfo{booktitle}{\emph{2019 IEEE/ACM
  41st International Conference on Software Engineering: Software Engineering
  Education and Training (ICSE-SEET)}}. IEEE, \bibinfo{pages}{121--125}.
\newblock


\bibitem[Clegg et~al\mbox{.}(2020)]%
        {clegg2020influence}
\bibfield{author}{\bibinfo{person}{Benjamin~S Clegg}, \bibinfo{person}{Phil
  McMinn}, {and} \bibinfo{person}{Gordon Fraser}.}
  \bibinfo{year}{2020}\natexlab{}.
\newblock \showarticletitle{The influence of test suite properties on automated
  grading of programming exercises}. In \bibinfo{booktitle}{\emph{2020 IEEE
  32nd Conference on Software Engineering Education and Training (CSEE\&T)}}.
  IEEE, \bibinfo{pages}{1--10}.
\newblock


\bibitem[Deiner et~al\mbox{.}(2023)]%
        {deiner2023automated}
\bibfield{author}{\bibinfo{person}{Adina Deiner}, \bibinfo{person}{Patric
  Feldmeier}, \bibinfo{person}{Gordon Fraser}, \bibinfo{person}{Sebastian
  Schweikl}, {and} \bibinfo{person}{Wengran Wang}.}
  \bibinfo{year}{2023}\natexlab{}.
\newblock \showarticletitle{Automated Test Generation for Scratch Programs}.
\newblock \bibinfo{journal}{\emph{Empirical Software Engineering}}
  \bibinfo{volume}{28}, \bibinfo{number}{3} (\bibinfo{year}{2023}).
\newblock
\urldef\tempurl%
\url{https://doi.org/10.1007/s10664-022-10255-x}
\showDOI{\tempurl}


\bibitem[Deiner and Fraser(2024)]%
        {deiner2024nuzzlebug}
\bibfield{author}{\bibinfo{person}{Adina Deiner} {and} \bibinfo{person}{Gordon
  Fraser}.} \bibinfo{year}{2024}\natexlab{}.
\newblock \showarticletitle{NuzzleBug: Debugging Block-Based Programs in
  Scratch}. In \bibinfo{booktitle}{\emph{Proceedings of the IEEE/ACM 46th
  International Conference on Software Engineering (ICSE'24)}}.
  \bibinfo{publisher}{ACM}, \bibinfo{numpages}{13}~pages.
\newblock
\urldef\tempurl%
\url{https://doi.org/10.1145/3597503.3623331}
\showDOI{\tempurl}


\bibitem[Edwards(2003)]%
        {edwards2003using}
\bibfield{author}{\bibinfo{person}{Stephen~H Edwards}.}
  \bibinfo{year}{2003}\natexlab{}.
\newblock \showarticletitle{Using test-driven development in the classroom:
  Providing students with automatic, concrete feedback on performance}. In
  \bibinfo{booktitle}{\emph{Proceedings of the international conference on
  education and information systems: technologies and applications EISTA}},
  Vol.~\bibinfo{volume}{3}. Citeseer.
\newblock


\bibitem[Fein et~al\mbox{.}(2022)]%
        {fein2022}
\bibfield{author}{\bibinfo{person}{Benedikt Fein}, \bibinfo{person}{Florian
  Oberm\"{u}ller}, {and} \bibinfo{person}{Gordon Fraser}.}
  \bibinfo{year}{2022}\natexlab{}.
\newblock \showarticletitle{CATNIP: An Automated Hint Generation Tool for
  Scratch}. In \bibinfo{booktitle}{\emph{Proceedings of the 27th ACM Conference
  on on Innovation and Technology in Computer Science Education Vol. 1}}
  (Dublin, Ireland) \emph{(\bibinfo{series}{ITiCSE '22})}.
  \bibinfo{publisher}{Association for Computing Machinery},
  \bibinfo{address}{New York, NY, USA}, \bibinfo{pages}{124–130}.
\newblock
\showISBNx{9781450392013}
\urldef\tempurl%
\url{https://doi.org/10.1145/3502718.3524820}
\showDOI{\tempurl}


\bibitem[Fraser et~al\mbox{.}(2021)]%
        {fraser2021litterbox}
\bibfield{author}{\bibinfo{person}{Gordon Fraser}, \bibinfo{person}{Ute Heuer},
  \bibinfo{person}{Nina K{\"o}rber}, \bibinfo{person}{Florian Oberm{\"u}ller},
  {and} \bibinfo{person}{Ewald Wasmeier}.} \bibinfo{year}{2021}\natexlab{}.
\newblock \showarticletitle{Litterbox: A Linter for Scratch Programs}. In
  \bibinfo{booktitle}{\emph{Proceedings of the International Conference on
  Software Engineering: Software Engineering Education and Training
  (ICSE-SEET'21)}}. \bibinfo{publisher}{IEEE}, \bibinfo{pages}{183--188}.
\newblock
\urldef\tempurl%
\url{https://doi.org/10.1109/ICSE-SEET52601.2021.00028}
\showDOI{\tempurl}


\bibitem[Frädrich et~al\mbox{.}(2020)]%
        {fraedrich2020}
\bibfield{author}{\bibinfo{person}{Christoph Frädrich},
  \bibinfo{person}{Florian Obermüller}, \bibinfo{person}{Nina Körber},
  \bibinfo{person}{Ute Heuer}, {and} \bibinfo{person}{Gordon Fraser}.}
  \bibinfo{year}{2020}\natexlab{}.
\newblock \showarticletitle{Common Bugs in Scratch Programs}. In
  \bibinfo{booktitle}{\emph{Proceedings of the 2020 ACM Conference on
  Innovation and Technology in Computer Science Education}} (Trondheim, Norway)
  \emph{(\bibinfo{series}{ITiCSE '20})}. \bibinfo{pages}{89--95}.
\newblock
\urldef\tempurl%
\url{https://doi.org/10.1145/3341525.3387389}
\showDOI{\tempurl}


\bibitem[Gruber et~al\mbox{.}(2021)]%
        {gruber2021empirical}
\bibfield{author}{\bibinfo{person}{Martin Gruber}, \bibinfo{person}{Stephan
  Lukasczyk}, \bibinfo{person}{Florian Kroi{\ss}}, {and}
  \bibinfo{person}{Gordon Fraser}.} \bibinfo{year}{2021}\natexlab{}.
\newblock \showarticletitle{An Empirical Study of Flaky Tests in Python}. In
  \bibinfo{booktitle}{\emph{Proceedings of the 14th IEEE Conference on Software
  Testing, Verification and Validation (ICST'21)}}. IEEE,
  \bibinfo{pages}{148--158}.
\newblock
\urldef\tempurl%
\url{https://doi.org/10.1109/ICST49551.2021.00026}
\showDOI{\tempurl}


\bibitem[Hovemeyer and Pugh(2004)]%
        {hovemeyer2004}
\bibfield{author}{\bibinfo{person}{David Hovemeyer} {and}
  \bibinfo{person}{William Pugh}.} \bibinfo{year}{2004}\natexlab{}.
\newblock \showarticletitle{Finding Bugs is Easy}.
\newblock \bibinfo{journal}{\emph{SIGPLAN Not.}} \bibinfo{volume}{39},
  \bibinfo{number}{12} (\bibinfo{date}{Dec.} \bibinfo{year}{2004}),
  \bibinfo{pages}{92–106}.
\newblock
\showISSN{0362-1340}
\urldef\tempurl%
\url{https://doi.org/10.1145/1052883.1052895}
\showDOI{\tempurl}


\bibitem[Johnson(2016)]%
        {johnson2016itch}
\bibfield{author}{\bibinfo{person}{David~E Johnson}.}
  \bibinfo{year}{2016}\natexlab{}.
\newblock \showarticletitle{ITCH: Individual Testing of Computer Homework for
  Scratch Assignments}. In \bibinfo{booktitle}{\emph{Proceedings of the 47th
  ACM Technical Symposium on Computing Science Education}}. ACM,
  \bibinfo{pages}{223--227}.
\newblock


\bibitem[Maloney et~al\mbox{.}(2010)]%
        {maloney2010}
\bibfield{author}{\bibinfo{person}{John Maloney}, \bibinfo{person}{Mitchel
  Resnick}, \bibinfo{person}{Natalie Rusk}, \bibinfo{person}{Brian Silverman},
  {and} \bibinfo{person}{Evelyn Eastmond}.} \bibinfo{year}{2010}\natexlab{}.
\newblock \showarticletitle{The Scratch Programming Language and Environment}.
\newblock \bibinfo{journal}{\emph{ACM Transactions on Computing Education
  (TOCE)}}  \bibinfo{volume}{10} (\bibinfo{date}{11} \bibinfo{year}{2010}),
  \bibinfo{pages}{16}.
\newblock
\urldef\tempurl%
\url{https://doi.org/10.1145/1868358.1868363}
\showDOI{\tempurl}


\bibitem[McGill and Decker(2020)]%
        {mcgill2020}
\bibfield{author}{\bibinfo{person}{Monica~M. McGill} {and}
  \bibinfo{person}{Adrienne Decker}.} \bibinfo{year}{2020}\natexlab{}.
\newblock \showarticletitle{Tools, Languages, and Environments Used in Primary
  and Secondary Computing Education}. In \bibinfo{booktitle}{\emph{Proceedings
  of the 2020 ACM Conference on Innovation and Technology in Computer Science
  Education}} (Trondheim, Norway) \emph{(\bibinfo{series}{ITiCSE '20})}.
  \bibinfo{publisher}{ACM}, \bibinfo{pages}{103–109}.
\newblock
\showISBNx{9781450368742}
\urldef\tempurl%
\url{https://doi.org/10.1145/3341525.3387365}
\showDOI{\tempurl}


\bibitem[Nurue and Gray(2024)]%
        {nurue2024}
\bibfield{author}{\bibinfo{person}{Herart~Dominggus Nurue} {and}
  \bibinfo{person}{Jeff Gray}.} \bibinfo{year}{2024}\natexlab{}.
\newblock \showarticletitle{A Testing Extension for Scratch}. In
  \bibinfo{booktitle}{\emph{Proceedings of the 2024 ACM Southeast Conference}}
  (Marietta, GA, USA) \emph{(\bibinfo{series}{ACMSE '24})}.
  \bibinfo{publisher}{Association for Computing Machinery},
  \bibinfo{address}{New York, NY, USA}, \bibinfo{pages}{266–271}.
\newblock
\showISBNx{9798400702372}
\urldef\tempurl%
\url{https://doi.org/10.1145/3603287.3651217}
\showDOI{\tempurl}


\bibitem[Oberm\"{u}ller et~al\mbox{.}(2023)]%
        {obermueller2023tutorials}
\bibfield{author}{\bibinfo{person}{Florian Oberm\"{u}ller},
  \bibinfo{person}{Luisa Greifenstein}, {and} \bibinfo{person}{Gordon Fraser}.}
  \bibinfo{year}{2023}\natexlab{}.
\newblock \showarticletitle{Effects of Automated Feedback in Scratch
  Programming Tutorials}. In \bibinfo{booktitle}{\emph{Proceedings of the 2023
  Conference on Innovation and Technology in Computer Science Education V. 1}}
  \emph{(\bibinfo{series}{ITiCSE 2023})}. \bibinfo{publisher}{Association for
  Computing Machinery}, \bibinfo{address}{New York, NY, USA},
  \bibinfo{pages}{396–402}.
\newblock
\showISBNx{9798400701382}
\urldef\tempurl%
\url{https://doi.org/10.1145/3587102.3588803}
\showDOI{\tempurl}


\bibitem[Oberm\"{u}ller et~al\mbox{.}(2021)]%
        {obermueller2021hinttest}
\bibfield{author}{\bibinfo{person}{Florian Oberm\"{u}ller},
  \bibinfo{person}{Ute Heuer}, {and} \bibinfo{person}{Gordon Fraser}.}
  \bibinfo{year}{2021}\natexlab{}.
\newblock \showarticletitle{Guiding Next-Step Hint Generation Using Automated
  Tests}. In \bibinfo{booktitle}{\emph{Proceedings of the 26th ACM Conference
  on Innovation and Technology in Computer Science Education V. 1}} (Virtual
  Event, Germany) \emph{(\bibinfo{series}{ITiCSE '21})}.
  \bibinfo{publisher}{Association for Computing Machinery},
  \bibinfo{address}{New York, NY, USA}, \bibinfo{pages}{220–226}.
\newblock
\showISBNx{9781450382144}
\urldef\tempurl%
\url{https://doi.org/10.1145/3430665.3456344}
\showDOI{\tempurl}


\bibitem[Panichella et~al\mbox{.}(2022)]%
        {panichella2022test}
\bibfield{author}{\bibinfo{person}{Annibale Panichella},
  \bibinfo{person}{Sebastiano Panichella}, \bibinfo{person}{Gordon Fraser},
  \bibinfo{person}{Anand~Ashok Sawant}, {and} \bibinfo{person}{Vincent~J
  Hellendoorn}.} \bibinfo{year}{2022}\natexlab{}.
\newblock \showarticletitle{Test smells 20 years later: detectability,
  validity, and reliability}.
\newblock \bibinfo{journal}{\emph{Empirical Software Engineering}}
  \bibinfo{volume}{27}, \bibinfo{number}{7} (\bibinfo{year}{2022}),
  \bibinfo{pages}{170}.
\newblock


\bibitem[Stahlbauer et~al\mbox{.}(2020)]%
        {stahlbauer2020}
\bibfield{author}{\bibinfo{person}{Andreas Stahlbauer},
  \bibinfo{person}{Christoph Frädrich}, {and} \bibinfo{person}{Gordon
  Fraser}.} \bibinfo{year}{2020}\natexlab{}.
\newblock \showarticletitle{{Verified from Scratch: Program Analysis for
  Learners’ Programs}}. In \bibinfo{booktitle}{\emph{In Proceedings of the
  International Conference on Automated Software Engineering (ASE)}}.
  \bibinfo{publisher}{{IEEE}}.
\newblock


\bibitem[Stahlbauer et~al\mbox{.}(2019)]%
        {stahlbauer2019testing}
\bibfield{author}{\bibinfo{person}{Andreas Stahlbauer}, \bibinfo{person}{Marvin
  Kreis}, {and} \bibinfo{person}{Gordon Fraser}.}
  \bibinfo{year}{2019}\natexlab{}.
\newblock \showarticletitle{Testing scratch programs automatically}. In
  \bibinfo{booktitle}{\emph{Proceedings of the 2019 27th ACM Joint Meeting on
  European Software Engineering Conference and Symposium on the Foundations of
  Software Engineering}}. \bibinfo{pages}{165--175}.
\newblock


\bibitem[Van~Deursen et~al\mbox{.}(2001)]%
        {van2001refactoring}
\bibfield{author}{\bibinfo{person}{Arie Van~Deursen}, \bibinfo{person}{Leon
  Moonen}, \bibinfo{person}{Alex Van Den~Bergh}, {and} \bibinfo{person}{Gerard
  Kok}.} \bibinfo{year}{2001}\natexlab{}.
\newblock \showarticletitle{Refactoring test code}. In
  \bibinfo{booktitle}{\emph{Proceedings of the 2nd international conference on
  extreme programming and flexible processes in software engineering
  (XP2001)}}. Citeseer, \bibinfo{pages}{92--95}.
\newblock


\bibitem[Voeten(2023)]%
        {poke}
\bibfield{author}{\bibinfo{person}{Iwijn Voeten}.}
  \bibinfo{year}{2023}\natexlab{}.
\newblock \showarticletitle{Een blokgebaseerd testframework voor Scratch}.
\newblock  (\bibinfo{year}{2023}).
\newblock
\newblock
\shownote{Master Thesis. \url{http://lib.ugent.be/catalog/rug01:003150096}.
  Scratch instance available at \url{https://scratch.ugent.be/poke/editor/}}.


\end{thebibliography}

\end{document}